\documentclass[twocolumn]{aastex63}

\usepackage{amsmath,amstext}
\usepackage[T1]{fontenc}
\usepackage{multirow}
\usepackage[utf8]{inputenc} 
\usepackage{float}
\usepackage{graphicx}
\usepackage[absolute,overlay]{textpos}
\usepackage{comment}
\usepackage{booktabs}

\newcommand{\unit}[1]{\ensuremath{\, \mathrm{#1}}}
\newcommand{\reshhe}{6\%}

\newcommand{\resincl}{$90\pm20^\circ$}

\newcommand{\PSUAA}{Department of Astronomy \& Astrophysics, 525 Davey Laboratory, The Pennsylvania State University, University Park, PA 16802, USA}
\newcommand{\PSUCEHW}{Center for Exoplanets and Habitable Worlds, 525 Davey Laboratory, The Pennsylvania State University, University Park, PA 16802, USA}

\newcommand{\Princeton}{Department of Astrophysical Sciences, Princeton University, 4 Ivy Lane, Princeton, NJ 08540, USA}
\newcommand{\Goddard}{NASA Goddard Space Flight Center, 8800 Greenbelt Road, Greenbelt, MD 20771, USA}

\newcommand{\MCDonald}{McDonald Observatory and Department of Astronomy, The University of Texas at Austin, 2515 Speedway, Austin, TX 78712, USA}
\newcommand{\UTSpace}{Center for Planetary Systems Habitability, The University of Texas at Austin, 2515 Speedway, Austin, TX 78712, USA}
\newcommand{\UCI}{Department of Physics \& Astronomy, The University of California, Irvine, Irvine, CA 92697, USA}

\newcommand{\Wyoming}{Department of Physics \& Astronomy, University of Wyoming, Laramie, WY 82070, USA}

\newcommand{\JPL}{Jet Propulsion Laboratory, California Institute of Technology, 4800 Oak Grove Drive, Pasadena, California 91109}
\newcommand{\HWS}{Department of Physics, Hobart and William Smith Colleges, 300 Pulteney Street, Geneva, NY 14456, USA}

\shorttitle{TOI-2015b: A Warm Neptune with Transit Timing Variations Orbiting an Active mid M Dwarf}
\shortauthors{Jones et al.}

\begin{document}
\title{TOI-2015b: A Warm Neptune with Transit Timing Variations Orbiting an Active mid M Dwarf}

\author[0000-0002-7227-2334]{Sinclaire E. Jones}
\affiliation{Department of Astronomy, The Ohio State University, 4055 McPherson Laboratory, Columbus, OH 43210, USA}
\affiliation{\Princeton}

\author[0000-0001-7409-5688]{Guðmundur Stefánsson}
\altaffiliation{NASA Sagan Fellow}
\affiliation{\Princeton}
\affiliation{Anton Pannekoek Institute for Astronomy, University of Amsterdam, Science Park 904, 1098 XH Amsterdam, The Netherlands} 

\author[0000-0003-1298-9699]{Kento Masuda}
\affil{Department of Earth and Space Science, Graduate School of Science, Osaka University, 1-1 Machikaneyama, Toyonaka, Osaka 560-0043, Japan}

\author[0000-0002-2990-7613]{Jessica E.\ Libby-Roberts}
\affil{\PSUAA}

\author[0000-0001-8621-6731]{Cristilyn N. Gardner}
\affil{\Wyoming}

\author[0000-0002-5034-9476]{Rae Holcomb}
\affil{\UCI}

\author[0000-0001-7708-2364]{Corey Beard}
\altaffiliation{NASA FINESST Fellow}
\affil{\UCI}

\author[0000-0003-0149-9678]{Paul Robertson}
\affil{\UCI}

\author[0000-0003-4835-0619]{Caleb I. Ca\~nas}
\altaffiliation{NASA Postdoctoral Fellow}
\affil{\Goddard}

\author[0000-0001-9596-7983]{Suvrath Mahadevan}
\affil{\PSUAA}
\affil{\PSUCEHW}

\author[0000-0001-8401-4300]{Shubham Kanodia}
\affil{Earth and Planets Laboratory, Carnegie Institution for Science, 5241 Broad Branch Road, NW, Washington, DC 20015, USA}

\author[0000-0002-9082-6337]{Andrea S.J.\ Lin}
\affil{\PSUAA}
\affil{\PSUCEHW}

\author[0000-0002-4475-4176]{Henry A. Kobulnicky}
\affil{\Wyoming}

\author[0000-0001-9307-8170]{Brock A. Parker}
\affil{\Wyoming}

\author[0000-0003-4384-7220]{Chad F. Bender}
\affil{Steward Observatory, The University of Arizona, 933 N. Cherry Avenue, Tucson, AZ 85721, USA}

\author[0000-0001-9662-3496]{William D. Cochran}
\affil{\MCDonald}
\affil{\UTSpace}

\author[0000-0002-2144-0764]{Scott A. Diddams}
\affil{Electrical, Computer \& Energy Engineering, University of Colorado, 425 UCB, Boulder, CO 80309, USA}
\affil{Department of Physics, University of Colorado, 2000 Colorado Avenue, Boulder, CO 80309, USA}
\affil{National Institute of Standards and Technology, 325 Broadway, Boulder, CO 80305, USA}

\author[0000-0002-3853-7327]{Rachel B. Fernandes}
\altaffiliation{President's Postdoctoral Fellow}
\affil{\PSUAA}
\affil{\PSUCEHW}

\author[0000-0002-5463-9980]{Arvind F. Gupta}
\affil{\PSUAA}
\affil{\PSUCEHW}

\author[0000-0003-1312-9391]{Samuel Halverson}
\affil{\JPL}

\author[0000-0002-6629-4182]{Suzanne L. Hawley}
\affil{Astronomy Department, University of Washington, Seattle, WA 98195}

\author[0000-0002-1664-3102]{Fred R. Hearty}
\affil{\PSUAA}
\affil{\PSUCEHW}

\author[0000-0003-1263-8637]{Leslie Hebb}
\affil{\HWS}
\affil{Department of Astronomy, Cornell University, 245 East Ave, Ithaca, NY 14850, USA}

\author[0000-0001-7458-1176]{Adam Kowalski}
\affil{Department of Astrophysical and Planetary Sciences, University of Colorado Boulder, 2000 Colorado Ave., Boulder, CO 80305, USA}
\affil{National Solar Observatory, University of Colorado Boulder, 3665 Discovery Drive, Boulder, CO 80303, USA}
\affil{Laboratory for Atmospheric and Space Physics, University of Colorado Boulder, 3665 Discovery Drive, Boulder, CO 80303, USA}

\author[0000-0001-8342-7736]{Jack Lubin}
\affil{\UCI}

\author[0000-0002-0048-2586]{Andrew Monson}
\affil{Steward Observatory, The University of Arizona, 933 N. Cherry Avenue, Tucson, AZ 85721, USA}

\author[0000-0001-8720-5612]{Joe P.\ Ninan}
\affil{Department of Astronomy and Astrophysics, Tata Institute of Fundamental Research, Homi Bhabha Road, Colaba, Mumbai 400005, India}

\author[0000-0002-4289-7958]{Lawrence Ramsey}
\affil{\PSUAA}
\affil{\PSUCEHW}

\author[0000-0001-8127-5775]{Arpita Roy}
\affil{Space Telescope Science Institute, 3700 San Martin Drive, Baltimore, MD 21218, USA}
\affil{Department of Physics and Astronomy, Johns Hopkins University, 3400 N Charles Street, Baltimore, MD 21218, USA}

\author[0000-0002-4046-987X]{Christian Schwab}
\affil{School of Mathematical and Physical Sciences, Macquarie University, Balaclava Road, North Ryde, NSW 2109, Australia}

\author[0000-0002-4788-8858]{Ryan C. Terrien}
\affil{Carleton College, One North College Street, Northfield, MN 55057, USA}

\author[0000-0001-9209-1808]{John Wisniewski}
\affil{NASA Headquarters, 300 Hidden Figures Way SW, Washington, D.C. 20546}

\begin{abstract}
We report the discovery of a close-in ($P_{\mathrm{orb}} = 3.349\:\mathrm{days}$) warm Neptune with clear transit timing variations (TTVs) orbiting the nearby ($d=47.3\:\mathrm{pc}$) active M4 star, TOI-2015. We characterize the planet’s properties using TESS photometry, precise near-infrared radial velocities (RV) with the Habitable-zone Planet Finder (HP) Spectrograph, ground-based photometry, and high-contrast imaging. A joint photometry and RV fit yields a radius $R_p~=~3.37_{-0.20}^{+0.15} \:\mathrm{R_\oplus}$, mass $m_p~=~16.4_{-4.1}^{+4.1}\:\mathrm{M_\oplus}$, and density $\rho_p~=~2.32_{-0.37}^{+0.38} \:\mathrm{g cm^{-3}}$ for TOI-2015b, suggesting a likely volatile-rich planet. The young, active host star has a rotation period of $P_{\mathrm{rot}}~=~8.7 \pm~0.9~\mathrm{days}$ and associated rotation-based age estimate of $1.1~\pm~0.1\:\mathrm{Gyr}$. Though no other transiting planets are seen in the TESS data, the system shows clear TTVs of super period $P_{\mathrm{sup}}~\approx~430\:\mathrm{days}$ and amplitude $\sim$$100\:\mathrm{minutes}$. After considering multiple likely period ratio models, we show an outer planet candidate near a 2:1 resonance can explain the observed TTVs while offering a dynamically stable solution. However, other possible two-planet solutions---including 3:2 and 4:3 resonance---cannot be conclusively excluded without further observations. Assuming a 2:1 resonance in the joint TTV-RV modeling suggests a mass of  $m_b~=~13.3_{-4.5}^{+4.7}\:\mathrm{M_\oplus}$ for TOI-2015b and $m_c~=~6.8_{-2.3}^{+3.5}\:\mathrm{M_\oplus}$ for the outer candidate. Additional transit and RV observations will be beneficial to explicitly identify the resonance and further characterize the properties of the system.
\end{abstract}


\section{Introduction}
\label{sec:intro}
M dwarfs are the most common type of star in the Milky Way galaxy and the lowest mass spectral type on the main sequence \citep{henry2006}. These low masses ($0.075~\unit{M_\odot}$ < $M_*$ < $0.6~\unit{M_\odot}$) make M dwarfs ideal hosts for planet detection and in-depth characterization. Based on results from the \textit{Kepler} mission, we know that M dwarfs tend to host more small planets on short period orbits than hotter, more massive stars \citep{dressing2015,muirhead2015,hardegree2019}. However, because the \textit{Kepler} mission observed a fixed field with a focus on FGK stars, it was only able to detect a limited number of M dwarf planets.

\begin{figure*}[!t]
\centering
\includegraphics[width=0.97\textwidth]{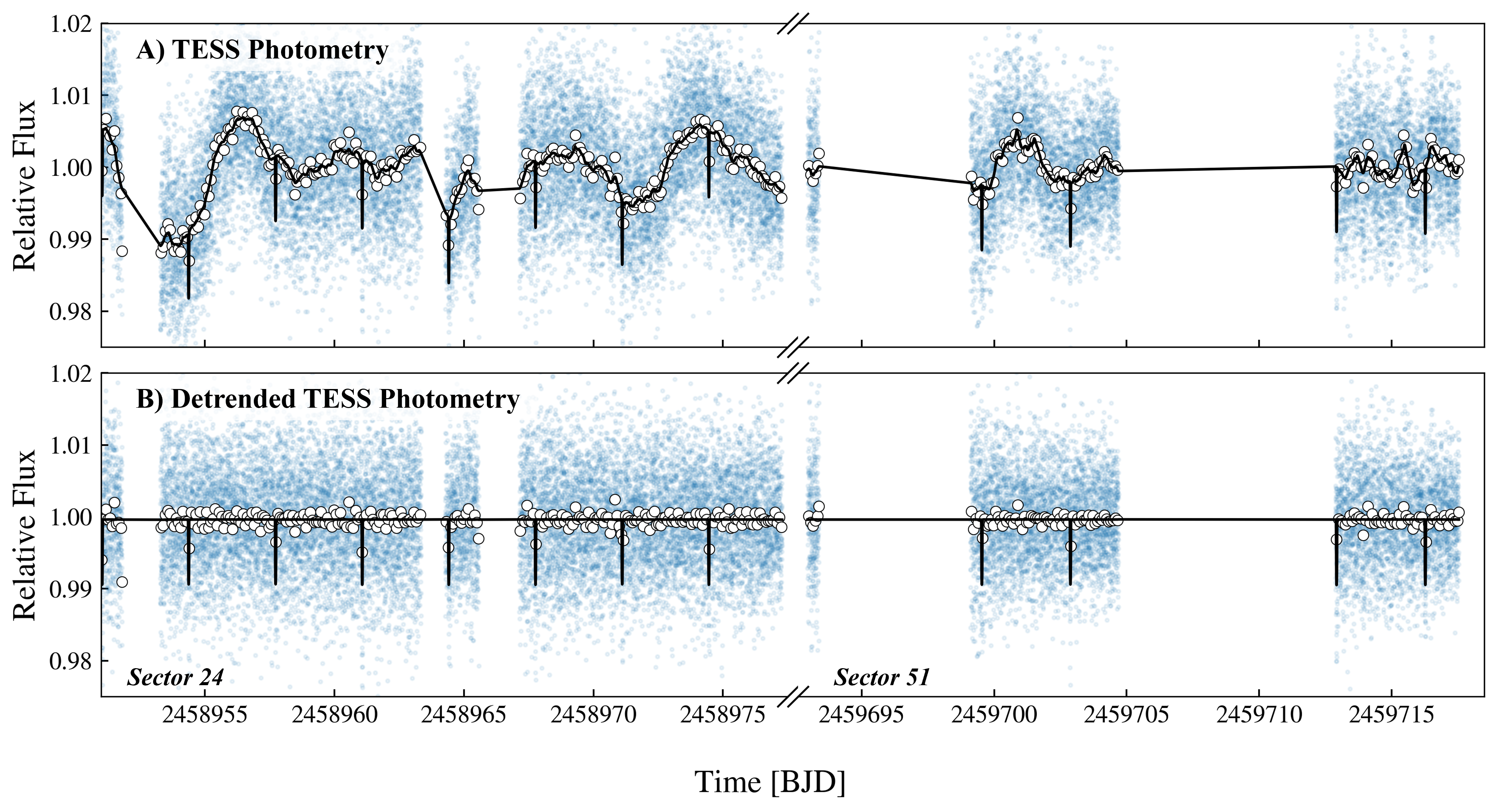}
\caption{TESS photometry of TOI-2015 from Sectors 24 and 51. Blue points are the PDCSAP data, and white points show 2hr binned data. \textit{A}) TESS photometry showing out-of-transit variability. The black curve shows the transit model along with the Gaussian Process quasi-periodic correlated noise model from our \texttt{juliet} fit (black curve). \textit{B}) TESS photometry after removing out-of-phase variability as modeled by the Gaussian Process. Transits of TOI-2015b are clearly visible.}
\vspace{0.3cm}
\label{fig:tess}
\end{figure*}

Both the \textit{K2} mission \citep{howell2014} and the Transiting Exoplanet Survey Satellite mission \citep[TESS;][]{ricker2015} have sampled a different population of planet hosts than \textit{Kepler}, including many nearby M dwarfs. In particular, with its red-optimized bandpass and all-sky coverage, TESS is particularly sensitive to planets orbiting nearby M dwarfs, having already detected hundreds of planetary candidates orbiting these cooler stars, which yield further insights into the population of planets orbiting nearby M stars \citep[e.g.,][]{gan2023,bryant2023,ment2023}.

Among the diverse demographics of exoplanets, multi-planet systems are especially interesting targets, as we can extract valuable clues to their formation from their orbital architectures. For instance, when planets are at or near orbital resonances, we may see evidence of transit timing variations \citep[TTVs;][]{holman2005,agol2005}---deviations from strictly periodic transits due to gravitational interactions between planets. Measured TTVs can be used to detect additional planets in the system and constrain their eccentricities and masses. It is theorized that TTV systems could have formed through convergent migration of planets while still embedded in a viscous protoplanetary disk \citep[e.g.,][]{masset2001,snellgrove2001,cresswell2006}.

Currently, there are about 150 confirmed TTV systems, only $\sim$10 of which have M dwarf hosts. Notable examples of M dwarf TTV systems include the well-studied TRAPPIST-1 system \citep{gillon2016, gillon2017}, a system of 7 transiting terrestrial planets in a resonant chain, and AU Mic, a young, active nearby M star with two transiting Neptunes \citep{plavchan2020,martioli2021}. There are also TTV systems that contain both transiting and non-transiting planets. For instance, K2-146 is a system of two resonant sub-Neptunes in which the initially non-transiting planet c precessed into view over time \citep{hirano2018b, hamann2019,lam2020}. Additionally, there is KOI-142, whose sub-Neptune has 12h TTVs \citep{nesvorny2013}. An outer planet near 2:1 resonance was later confirmed in this system via radial velocity (RV) follow-up observations \citep{barros2014}. 

Here we report on the discovery of a close-in ($P_{\mathrm{orb}} = 3.349 \unit{days}$) warm Neptune with clear TTVs transiting the nearby ($d=47.3\unit{pc}$) active mid M dwarf, TOI-2015. We confirm the planetary nature of the transiting object using TESS photometry along with ground-based photometric observations, high-contrast imaging, and precise RV observations. Photometry from TESS and ground-based instruments displays clear evidence of TTVs with a super-period of $P_{\mathrm{sup}}~\approx~430~\unit{days}$ and an amplitude of $\sim~100~\unit{min}$. However, the two sectors of TESS data reveal no significant evidence of additional transiting planets in the system. We constrain the mass of TOI-2015b using precise near-infrared (NIR) RVs obtained with the Habitable-zone Planet Finder \citep[HPF;][]{mahadevan2012, mahadevan2014} on the 10m Hobby-Eberly Telescope along with a joint TTV and RV fit assuming likely period ratios of the system.

This paper is organized as follows. Section \ref{sec:obs} describes the observations and data reduction. In Section \ref{sec:stellarparams}, we report the key parameters of the host star. Section \ref{sec:planetparams} provides an in-depth look at the transit, radial velocity, and TTV modeling and the resulting planet parameter constraints. In Section \ref{sec:discussion}, we place TOI-2015 in context with other M dwarfs and multi-planet systems with detectable TTVs. We conclude in Section \ref{sec:summary} with a summary of our key findings.

\section{Observations and Data Reduction}
\label{sec:obs}

\subsection{TESS Photometry}
TOI-2015 is listed as TIC 368287008 in the TESS Input Catalog \citep{stassun2018, stassun2019}, and is included in the mission's catalog of cool dwarf targets \citep{muirhead2018}. TESS observed TOI-2015 with a 2 minute cadence in two sectors: Sector 24 from 2020 April 16 to 2020 May 13 and Sector 51 from 2022 April 22 to 2022 May 18.  Analysis of the light curve by the TESS Science Processing Operations Center (SPOC) identified a possible planetary signal, TOI-2015.01 (available on the TESS alerts website\footnote{\url{https://tev.mit.edu/data/}}), where SPOC data validation reports note no significant centroid offsets during transit events \citep{twicken2018, li2019}.

\begin{figure*}[ht!]
\centering
\includegraphics[width=0.8\textwidth]{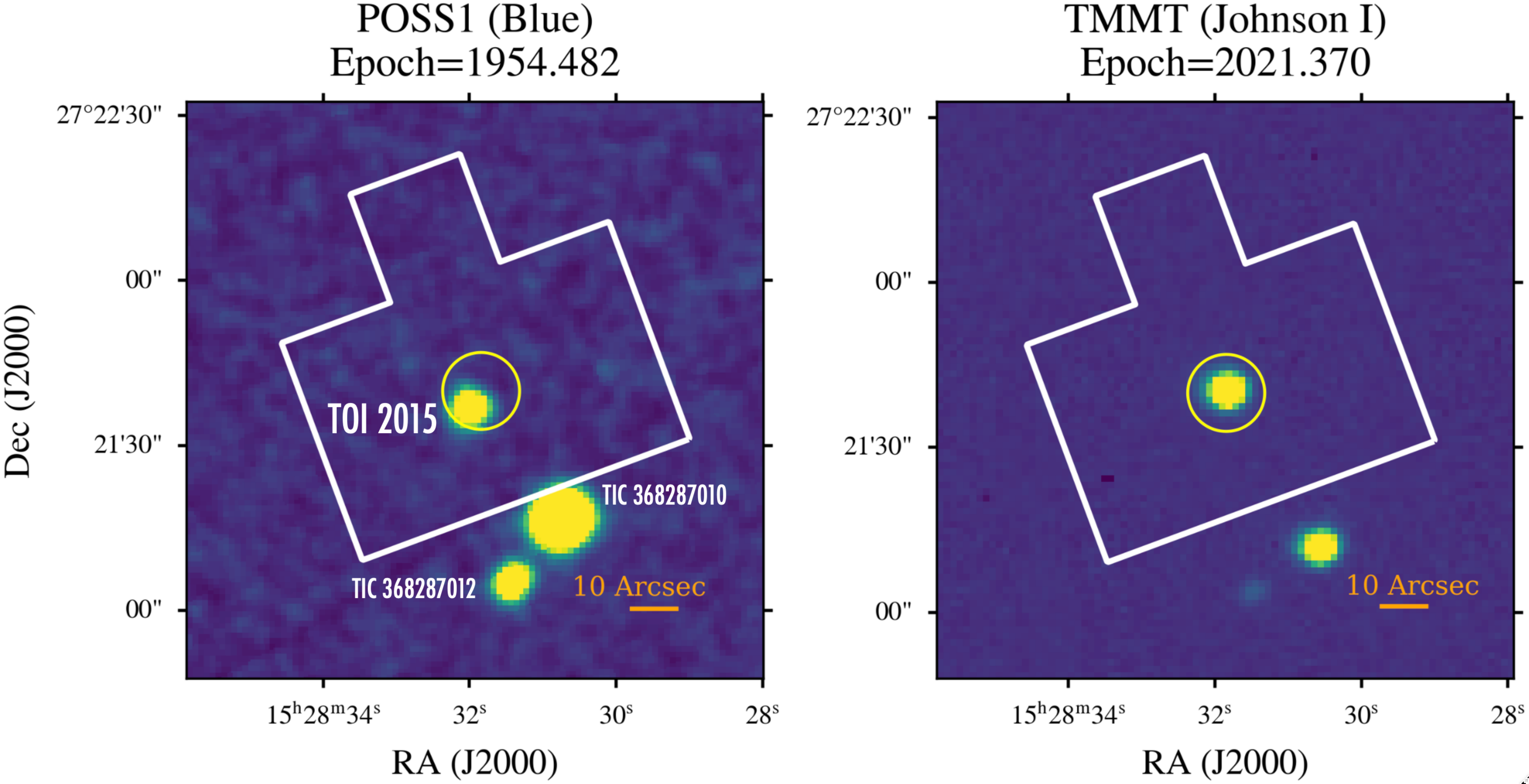}
\caption{Seeing limited imaging. (Left): Seeing limited imaging from POSS-1 in the Blue filter from the Digital Sky Survey-1 (DSS-1) from 1954. (Right): Seeing limited imaging obtained with TMMT in the Johnson~I filter in May 2021. The filled yellow circle ($7\arcsec$ radius) denotes the Gaia position of TOI-2015 at the Gaia epoch (2015.5). The white lines show the outline of the TESS aperture. North is up, and East is to the left. The two nearby stars TIC 368287010 and TIC 368287012 result in a modest dilution of $0.131892 \pm 0.001231$ in the TESS aperture. No other nearby stars are detected.}
\vspace{0.3cm}
\label{fig:seeing}
\end{figure*}

The TOI-2015 TESS photometry is displayed in Figure~\ref{fig:tess}. We retrieved these data using the \texttt{lightkurve} package \citep{lightkurve}. For our photometric analysis, we used the Presearch Data Conditioning Single Aperture Photometry (PDCSAP) light curve, which uses pixels chosen to maximize the signal-to-noise ratio (S/N) of the target and has removed systematic variability by fitting out trends common to many stars \citep{smith2012,stumpe2014}. Figure~\ref{fig:seeing} highlights the TESS aperture of TOI-2015 (Tmag=12.8), along with two known nearby objects within $40\arcsec$ as detected by Gaia: TIC 368287010 (Tmag=13.0; angular separation=32$\arcsec$) and TIC 368287012 (Tmag=16.5; angular separation=37$\arcsec$). From their separation and lower brightness than TOI-2015, the nearby stars result in a modest dilution of the TESS light curve with a contamination ratio of $0.131892 \pm 0.001231$. The flux contamination, which can be computed using the \texttt{tic\_contam.py} script\footnote{\url{https://github.com/mpaegert/tic_inspect/blob/master/tic_contam.py}} from \cite{paegert2021}, is corrected for in the PDCSAP light curves as prepared by SPOC.

\subsection{Seeing Limited Imaging}
To constrain blends within the TESS aperture, Figure~\ref{fig:seeing} compares seeing limited images of TOI-2015 observed in 1954 and 2021. The 1954 image was captured during the first Palomar Sky Survey (POSS1), and we accessed these data with \texttt{astroquery skyview} \citep{astroquery}. In the more recent seeing limited image, we observed TOI-2015 with the Three-hundred MilliMeter Telescope \citep[TMMT; ][]{monson2017} at Las Campanas Observatory on 2021 May 16. We obtained the TMMT image with the Johnson~I filter and an exposure time of 120 seconds. The nearby stars, TIC 368287010 (Tmag=13.0) and TIC 368287012 (Tmag=16.5), are highlighted in Figure~\ref{fig:seeing}. Due to TOI-2015's modest proper motion of $\mu_\alpha = -56~\unit{mas\:yr^{-1}}$ and $\mu_\delta = 64~\unit{mas\:yr^{-1}}$, TOI-2015 has moved slightly from the POSS1 epoch to the 2021 epoch. No background star is seen in the POSS1 data at the current location of TOI-2015 that could be a significant source of dilution.

\subsection{High-Contrast Imaging with WIYN 3.5m/NESSI and Lick 3m/ShaneAO}
\begin{figure*}[htp!]
\centering
\includegraphics[width=0.97\textwidth]{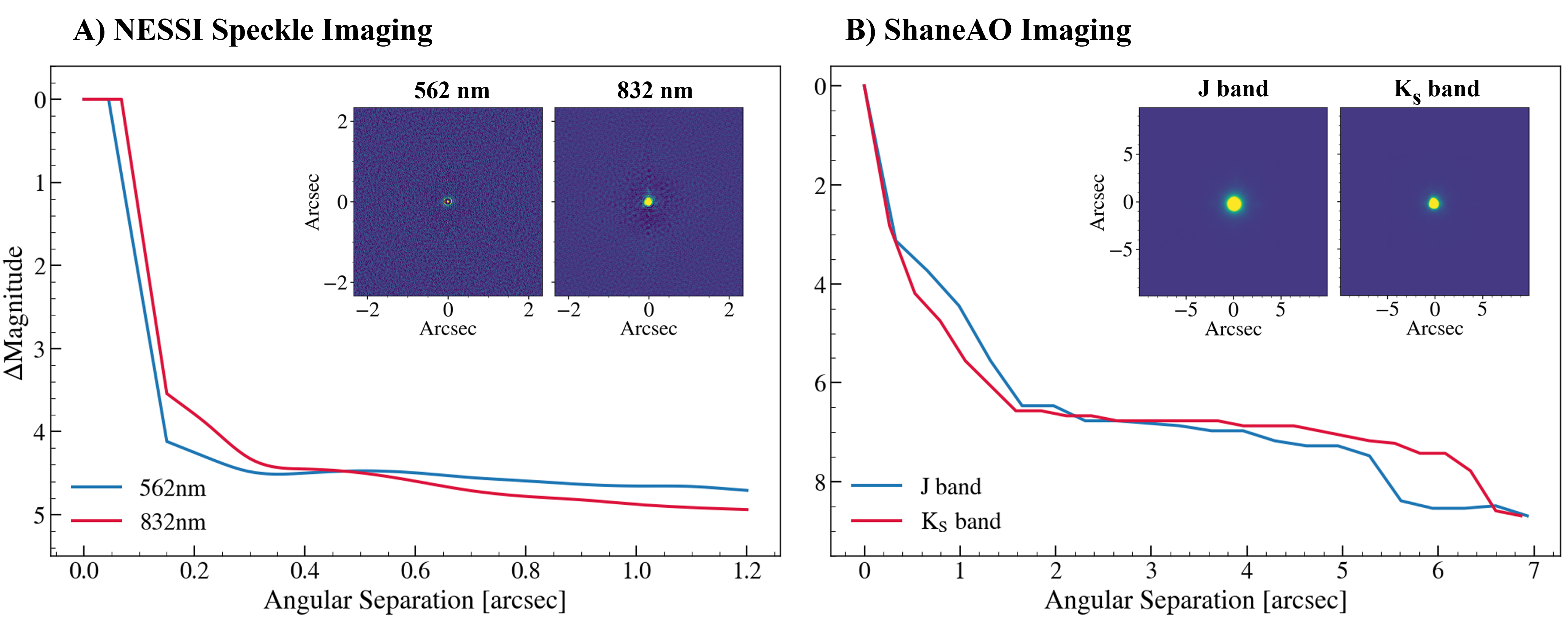}
\caption{High-contrast imaging of TOI-2015 from \textit{A}) WYIN 3.5m/NESSI Speckle imaging, and \textit{B}) Lick 3m/ShaneAO. The curves show azimuthally averaged radial $\Delta$ magnitude constraints, in the 562nm (blue) and 832nm (red) filters for NESSI, and the $J$ and $K_S$ band for ShaneAO. The inset images show the corresponding high-contrast images. We see no evidence of nearby companions in either the NESSI or the ShaneAO observations.}
\label{fig:imaging}
\end{figure*}

We obtained speckle imaging of TOI-2015 on 2021 March 29 with the NASA Exoplanet Star and Speckle Imager \citep[NESSI;][]{scott2018} on the WIYN 3.5m telescope to determine if the transit signals could be explained by contamination from nearby stars and other false positives (e.g., background eclipsing binary).\footnote{The WIYN Observatory is a joint facility of the University of Wisconsin-Madison, Indiana University, Purdue University, Penn State University, Princeton University, NSF's National Optical-Infrared Astronomy Research Laboratory, and NASA.} The target was observed in narrowband filters centered around 562nm and 832nm. The data were reduced using the standard NESSI pipeline \citep{howell2011}. The resulting contrast curves and their associated $4.6~\arcsec \times 4.6~\arcsec$ images are displayed in Figure~\ref{fig:imaging}. We place a $\Delta\mathrm{mag}\sim4.5$ limit on nearby objects between an angular separation of $0.2~\arcsec$ to $1.2~\arcsec$ and identify no additional nearby sources.

Additionally, we observed TOI-2015 with high-contrast adaptive optics (AO) imaging with the ShaneAO system on the 3m Telescope at Lick Observatory on 2021 May 27 \citep{gavel2014}. This allows us to further discern potential transit false positives. We observed in the $K_s$ and $J$ bands with 60s and 150s exposures respectively. The data were reduced following \cite{stefansson2020}. These results are shown in Figure~\ref{fig:imaging}. We identify no nearby companions of $\Delta\mathrm{mag}\sim7$ within a $1.5~\arcsec$ to $7~\arcsec$ radius of TOI-2015.

\subsection{Near-infrared RVs with HET 10m/HPF}
The Habitable-zone Planet Finder \citep[HPF;][]{mahadevan2012, mahadevan2014} is a near-infrared fiber-fed \citep{kanodia2018fiber} spectrograph on the 10-meter Hobby-Eberly Telescope (HET) at McDonald Observatory in Texas. HPF covers the information rich $z$, $Y$, and $J$ bands ($810-1280~\unit{nm}$) with a spectral resolution of R~$\sim55,000$. In order to achieve precise radial velocity measurements, the instrument's optics are kept under high-quality vacuum at a constant operating temperature of $\sim180~\unit{K}$ with milli-Kelvin long-term stability \citep{stefansson2016}. All observations were executed within the HET queue \citep{shetrone2007}. HPF has a laser frequency comb (LFC) calibrator that can provide $\sim20~\unit{cm/s}$ RV calibration precision in $\sim10~\unit{min}$ bins \citep{metcalf2019}. Due to the faintness of the target, we followed \cite{stefansson2020}; we did not use the simultaneous LFC in order to minimize the risk of contaminating the science spectrum from scattered light, and we extrapolated the wavelength solution from LFC calibration exposures that were taken throughout the night. This methodology has been shown to be precise at the $\sim30~\unit{cm/s}$ level, substantially smaller than the median photon-limited RV precision we obtain on TOI-2015.

In total, we obtained 109 HPF spectra in over 40 HET visits with 650 second exposures and a median S/N of $\sim45$ evaluated per 1D extracted pixel at $1~\unit{\mu m}$. There were 5 spectra that had a S/N < 17, which we removed from the analysis. The remaining 104 spectra were obtained in 37 HET visits, have a median S/N of 46, and span a baseline of 639 days. The spectra have a median unbinned RV uncertainty of $24~\unit{m/s}$ and a nightly (2-3 visits per night) binned RV uncertainty of $15~\unit{m/s}$. We used the binned measurements for our RV analysis.

We extracted the HPF 1D spectra using the HPF pipeline, following the procedures in \cite{ninan2018}, \cite{kaplan2018}, and \cite{metcalf2019}. We extracted high-precision HPF RVs using a modified version of the spectrum radial velocity analyzer \citep[\texttt{SERVAL};][]{zechmeister2018} pipeline that we have optimized to extract RVs for HPF spectra, following the procedures in \cite{stefansson2020}. \texttt{SERVAL} uses the template-matching technique to extract precise RVs for M dwarfs. To calculate barycentric corrections for the HPF spectra, we use the \texttt{barycorrpy} package \citep{kanodia2018}, which uses the methodology of \cite{wright2014} to calculate barycentric velocities. We use the 10 HPF orders least affected by tellurics, covering the wavelength regions from $8540-8890~\unit{\textup{\AA}}$, $9940-10940~\unit{\textup{\AA}}$, and $12375-12525~\unit{\textup{\AA}}$. We subtracted the estimated sky-background from the stellar spectrum using the dedicated HPF sky fiber. Following the methodology described in \cite{metcalf2019} and \cite{stefansson2020}, we explicitly masked out telluric lines and sky-emission lines to minimize their impact on the RV determination. Table \ref{tab:rvs} in the Appendix lists the RVs from HPF used in this work.

\subsection{Ground-based Follow-up Photometry}

\subsubsection{LCOGT 1m Photometry}
A partial transit of TOI-2015b was observed by the Las Cumbres Observatory Global Telescope (LCOGT) observing team on the night of 2020 July 4 using the Sinistro imaging cameras on the 1m telescope at its South African Astronomical Observatory site \citep{brown2013}. The Bessell~I filter was used with an exposure time of $100~\unit{s}$. The imager uses $1\times1$ binning in its full frame configuration with a gain of $10.3~\unit{e^-/ADU}$ and read noise of  $7.6~\unit{e^-}$. The camera has a plate scale of $0.389~\unit{\arcsec/pix}$ with a field of view of $26.5~\arcmin \times 26.5~\arcmin$. We accessed the data, processed using the BANZAI pipeline \citep{curtis_mccully_2018_1257560}, through the publicly accessible LCOGT archive (Proposal ID: KEY2020B-005, PI: Shporer).\footnote{\url{https://archive.lco.global/}}

We reduced the photometry using \texttt{AstroImageJ} \citep{collins2017}, which accounts for photometric errors due to photon, read, dark, digitization, and background noise. In addition, we added the expected error due to scintillation following the methodology in \cite{stefansson2017}. After testing a number of different aperture sizes and reference stars, we converged on a final aperture radius of $10~\unit{pix}$ with an inner sky background annulus of $16~\unit{pix}$ and outer annulus of $24~\unit{pix}$, as this extraction provided the highest overall photometric precision. 

\subsubsection{Diffuser-assisted WIRO 2.3m Photometry}
We obtained a full transit of TOI-2015b on the night of 2021 July 18 with the Wyoming Infrared Observatory (WIRO) DoublePrime prime-focus imager on the WIRO 2.3\,m Telescope \citep{Findlay2016}. The images used the SDSS~$i^\prime$ filter and an exposure time of $75~\unit{s}$. In the $1\times1$ binning mode, the WIRO detector has a gain of $2.6~\unit{e^-/ADU}$, read noise of $5.2~\unit{e^-}$. The four-amplifier mode has a $20~\unit{s}$ readout time. The plate scale is $0.576~\unit{\arcsec/pix}$ covering a field of view of $39~\arcmin \times 39~\arcmin$.

We used the Engineered Diffuser---a nanofabricated piece of optic able to mold the image of a star into a broad and stabilized shape, which can help provide high-precision photometry \citep{stefansson2017,stefansson2018a,stefansson2018b}---available on the WIRO DoublePrime imager \citep{Gardner-Watkins2023}. 

We used \texttt{AstroImageJ} to extract the data, and added the expected scintillation noise errors following \cite{stefansson2017}. After testing a number of aperture settings and reference stars, we found that an aperture radius of $25~\unit{pix}$, inner sky background annulus of $100~\unit{pix}$, and outer annulus of $150~\unit{pix}$ yielded the highest precision extraction. We adopt this extraction for our analysis.

\subsubsection{RBO 0.6m Photometry}
On the night of 2023 April 23, we observed a full TOI-2015b transit using the Apogee Alta F16 camera on the 0.6m telescope at Red Buttes Observatory \citep[RBO;][]{kasper2016} in Wyoming. We used the Bessell~I filter and 240 second exposures. In the $2\times2$ binning mode, the detector has a gain of $1.4~\unit{e^-/ADU}$, and read noise of $16~\unit{e^-}$. The plate scale is $0.72~\unit{\arcsec/pix}$ with a $25\arcmin \times 25\arcmin$ field of view.

We extracted the photometry using a custom \texttt{python} pipeline adapted from the one outlined in \cite{monson2017}. After testing a number of reference star and aperture combinations, we chose to use a $6~\unit{pix}$ aperture with an inner sky radius of $20~\unit{pix}$ and outer sky radius of $40~\unit{pix}$.

\subsubsection{ARC 3.5m Photometry}
We obtained a full transit of TOI-2015b on the night of 2023 April 23 with the Astrophysical Research Consortium Telescope Imaging Camera \citep[ARCTIC;][]{huehnerhoff2016} on the Astrophysical Research Council (ARC) 3.5m telescope at the Apache Point Observatory. Due to weather, we chose to observe with slight defocusing in the narrowband Semrock filter \citep[8570/300\AA;][]{stefansson2017,stefansson2018b}, which avoids atmospheric absorption lines, using 40 second exposures. The data were reduced using \texttt{AstroImageJ}. After experimenting with a number of apertures and reference stars, we chose to use a $16~\unit{pix}$ aperture with a $25~\unit{pix}$ inner sky background annulus and $35~\unit{pix}$ outer annulus.

On the night of 2023 May 3, we observed another full transit of TOI-2015b. We used the SDSS~$i^\prime$ filter and 45 second exposures with the Engineered Diffuser available on ARCTIC \citep{stefansson2017}. We used the \texttt{astroscrappy}\footnote{\texttt{\url{https://github.com/astropy/astroscrappy}}} code to correct for cosmic ray hits and reduced the photometry with \texttt{AstroImageJ} using an aperture of $12~\unit{pix}$ and inner and outer sky annulus of 20 and $40~\unit{pix}$ respectively for our final reduction.

During both nights, we used $4\times4$ binning with a gain of $1.9~\unit{e^-/ADU}$ and read noise of $6.6~\unit{e^-}$. We also used the quad amplifier with the $2.7~\unit{sec}$ fast readout rate mode. In this binning mode, ARCTIC has a plate scale of $0.456~\unit{\arcsec/pix}$, and a $7.9~\arcmin \times 7.9~\arcmin$ field of view.

\begin{deluxetable*}{llcc}
\tablecaption{Summary of TOI-2015 stellar parameters used in this work.}
\tabletypesize{\scriptsize}
\tablehead{\colhead{~~~Parameter}   &  \colhead{Description}                                          & \colhead{Value}               & \colhead{Reference}}
\startdata
\multicolumn{4}{l}{\hspace{-0.3cm} \textbf{Equatorial Coordinates, Proper Motion, and Spectral Type:}}                                                                                         \\
$\alpha_{\mathrm{J2000}}$          & Right Ascension (RA), J2015.5                                    & 15:28:31.84                   & Gaia DR3                          \\
$\delta_{\mathrm{J2000}}$          & Declination (Dec), J2015.5                                       & +27:21:39.86                  & Gaia DR3                          \\
$\mu_{\alpha}$                     & Proper motion (RA, \unit{mas\:yr^{-1}})                           & $-56.244 \pm 0.010$           & Gaia DR3                          \\
$\mu_{\delta}$                     & Proper motion (Dec, \unit{mas\:yr^{-1}})                          & $63.807 \pm 0.014$            & Gaia DR3                          \\
\multicolumn{4}{l}{\hspace{-0.3cm} \textbf{Equatorial Coordinates, Proper Motion, and Spectral Type:}}                                                                                         \\
$B$                                & APASS Johnson B mag                                              & $17.134 \pm 0.102$            & APASS                         \\
$V$                                & APASS Johnson V mag                                              & $16.11 \pm 0.2$               & APASS                         \\
$g^{\prime}$                       & APASS Sloan $g^{\prime}$ mag                                     & $16.284 \pm 0.018$            & APASS                         \\
$r^{\prime}$                       & APASS Sloan $r^{\prime}$ mag                                     & $14.988 \pm 0.019$            & APASS                         \\
$i^{\prime}$                       & APASS Sloan $i^{\prime}$ mag                                     & $13.358 \pm 0.017$            & APASS                         \\
\textit{TESS}-mag & \textit{TESS} magnitude                                                           & $12.8387 \pm 0.007$           & TIC                           \\
$J$                                & 2MASS $J$ mag                                                    & $11.118 \pm 0.022$            & 2MASS                         \\
$H$                                & 2MASS $H$ mag                                                    & $10.519 \pm 0.019$            & 2MASS                         \\
$K_S$                              & 2MASS $K_S$ mag                                                  & $10.263 \pm 0.019$            & 2MASS                         \\
$WISE1$                            & WISE1 mag                                                        & $10.098 \pm 0.022$            & WISE                          \\
$WISE2$                            & WISE2 mag                                                        & $9.926 \pm 0.02$              & WISE                          \\
$WISE3$                            & WISE3 mag                                                        & $9.771 \pm 0.039$             & WISE                          \\
$WISE4$                            & WISE4 mag                                                        & $9.319$                       & WISE                          \\
\multicolumn{4}{l}{\hspace{-0.3cm} \textbf{Spectroscopic Parameters from \texttt{HPF-SpecMatch}:}}                                                                                              \\
$T_{\mathrm{eff}}$                 & Effective temperature in \unit{K}                                & $3194 \pm 56$                 & This work                     \\ 
$\mathrm{[Fe/H]}$                  & Metallicity in dex                                               & $0.05 \pm 0.14$               & This work                     \\ 
$\log(g)$                          & Surface gravity in cgs units                                     & $4.95 \pm 0.04$               & This work                     \\ 
\multicolumn{4}{l}{\hspace{-0.3cm} \textbf{Model-Dependent Stellar SED and Isochrone fit Parameters$^a$:}}                                                                           \\
Spectral Type                      & -                                                                & M4                            & This Work                     \\
$M_*$                              & Mass in $M_{\odot}$                                              & $0.342^{+0.021}_{-0.025}$     & This work                     \\
$R_*$                              & Radius in $R_{\odot}$                                            & $0.333 \pm 0.011$             & This work                     \\
$\rho_*$                           & Density in $\unit{g\:cm^{-3}}$                                   & $12.98^{+1.1}_{-0.97}$        & This work                     \\
Age                                & Age in Gyr from SED fit                                         & $6.8^{+4.7}_{-4.6}$           & This work                     \\
$L_*$                              & Luminosity in $L_\odot$                                          & $0.01050^{+0.00044}_{-0.00052}$& This work                     \\
$d$                                & Distance in pc                                                   & $47.32_{-0.03}^{+0.04}$       & Gaia DR3, Bailer-Jones            \\
$\pi$                              & Parallax in mas                                                  & $21.131 \pm 0.018$            & Gaia DR3                          \\
\multicolumn{4}{l}{\hspace{-0.3cm} \textbf{Other Stellar Parameters:}}                                                                                                                         \\
$v \sin i_\star$                   & Stellar rotational velocity in $\unit{km\ s^{-1}}$               & $3.2\pm0.6$                   & This work                     \\
$P_{\mathrm{rot}}$                 & Rotation Period in days                                          & $8.7\pm0.9$                   & This work                     \\
Age                                & Gyrochronological Age in Gyr$^b$                                & $1.1\pm 0.1$                  & This work                     \\
$i_\star$                          & Stellar inclination in degrees                                   & $90\pm 20^{\circ}$            & This work                     \\ 
$RV$                               & Absolute radial velocity in $\unit{km\ s^{-1}}$ ($\gamma$)       & $-32.04\pm0.17$               & This work                     \\
$U$                                & Galactic $U$ Velocity (km/s)                                     & $-30.60\pm0.07$               & This work                     \\
$V$                                & Galactic $V$ Velocity (km/s)                                     & $-11.51\pm0.07$               & This work                     \\
$W$                                & Galactic $W$ Velocity (km/s)                                     & $-17.95\pm0.14$               & This work                     \\
\enddata
\tablenotetext{}{References are: TIC \citep{stassun2018,stassun2019}, Gaia DR3 \citep{gaia_collaboration_gaia_2023}, APASS \citep{henden2015apass}, 2MASS/WISE \citep{cutri2014wise}, Bailer-Jones \citep{bailer-jones2021}.}
\tablenotetext{a}{{\tt EXOFASTv2} derived values using MIST isochrones with the Gaia parallax and spectroscopic parameters from $a$) as priors.}
\tablenotetext{b}{Using the gyrochronological age relation for M4-M6.5 dwarfs in \cite{engle2023}. Uncertainties are likely underestimated.}
\label{tab:stellarparam}
\end{deluxetable*}

\section{Stellar Parameters}
\label{sec:stellarparams}

\subsection{Spectroscopic Parameters}
To obtain spectroscopic constraints on the effective temperature ($T_{\mathrm{eff}}$), metallicity ([Fe/H]), surface gravity ($\log~g$), and projected rotational velocity ($v~\sin~i_\star$) of TOI-2015, we used the \texttt{HPF-SpecMatch}\footnote{\texttt{\url{https://github.com/gummiks/hpfspecmatch}}} code \citep{stefansson2020}. This code uses a two-step $\chi^2$ minimization process to compare a target spectrum to an as-observed spectral library of 166 well-characterized stars ($2700 \unit{K}$ < $T_{\mathrm{eff}}$ < $5990 \unit{K}$). The final target parameters are determined with a weighted average of the five best-fitting library star parameters.

Using \texttt{HPF-SpecMatch}, we determine the following stellar parameters for TOI-2015: $T_\mathrm{{eff}} = 3194\pm 56~\unit{K}$, log~$g = 4.95\pm 0.04$, and [Fe/H] = $0.05\pm 0.14$, where the uncertainties are calculated using a leave-one-out cross-validation process on the full spectral library. For the projected rotational velocity, we obtain $v~\sin~i_\star = 3.2 \pm 0.6~\unit{km/s}$, where the uncertainty adopted is the standard deviation of the $v~\sin~i_\star$ values from running our spectral matching analysis on the 7 orders cleanest of tellurics (8540-8640\AA, 8670-8750\AA, 8790-8885\AA, 9940-10055\AA, 10105-10220\AA, 10280-10395\AA, 10460-10570\AA). Overall, the two best fitting stars are GJ 1289 and GJ 4065, which have spectral types of M4.5 \citep{lepine2013}, and M4 \citep{newton2014}, respectively. As both are in general agreement, we adopt a spectral type of M4 for TOI-2015.

\subsection{Spectral Energy Distribution}
\label{sec:sed}
We fit the Spectral Energy Distribution (SED) of TOI-2015 using the exoplanet and stellar fitting software package \texttt{EXOFASTv2} \citep{eastman2019}. This allows us to constrain values such as mass, radius, and age for TOI-2015. The SED fit requires three primary inputs: available literature photometry data, parallax given by Gaia \citep{bailer-jones2021}, and the spectroscopic parameters derived using \texttt{HPF-SpecMatch}. As the model dependent SED parameters for $T_\mathrm{{eff}}$, $\log g$, and [Fe/H] agree with the spectroscopic parameters we obtain with \texttt{HPF-SpecMatch}, we do not list them in Table \ref{tab:stellarparam} for simplicity.

\begin{figure*}[hbt!]
\centering
\includegraphics[width=0.9\textwidth]{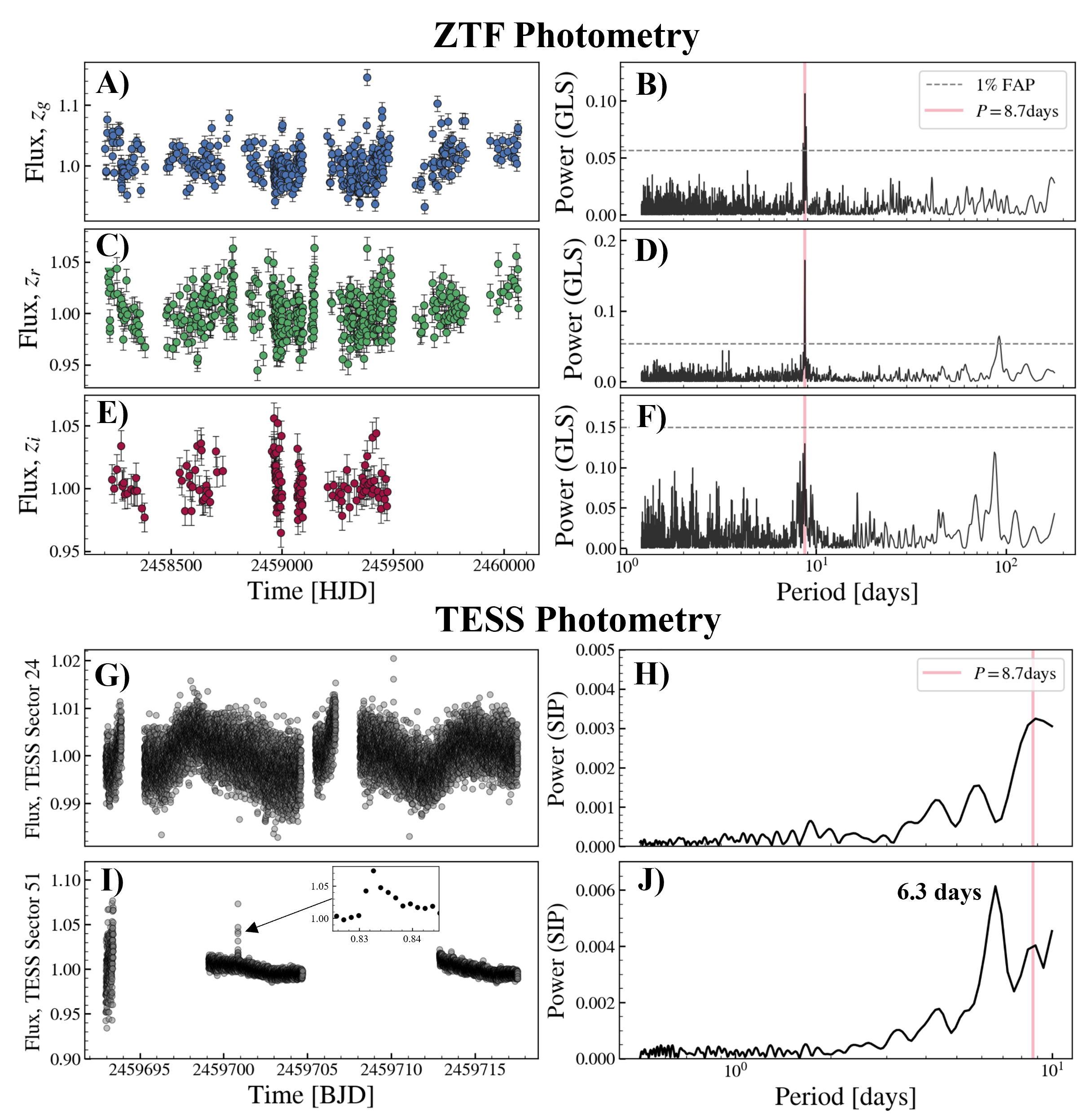}
\caption{Out-of-transit photometry from ZTF and TESS. \textit{A-F}) ZTF photometry and associated Generalized Lomb-Scargle periodograms in the $z_g$, $z_r$, and $z_i$ filters. All filters show a peak at 8.7 days (red vertical line). \textit{G-J}) TESS photometry and associated TESS Systematics-Insensitive Periodogram (TESS-SIP) plots for Sectors 24 and 51. TESS Sector 24 shows a peak consistent with 8.7 days, whereas Sector 51 shows a peak at 6.3 days. We attribute the latter to low data coverage and large gaps in the TESS data. We adopt a rotation period of $8.7\pm0.9 \unit{days}$. In panel \textit{I}, we also show a zoomed in view of one of the flares in the TESS photometry.}
\label{fig:ztf}
\end{figure*}

\subsection{TOI-2015 is a fully-convective star}\label{sec:jaogap}
\begin{figure}[t!]
\centering
\includegraphics[width=0.8\columnwidth]{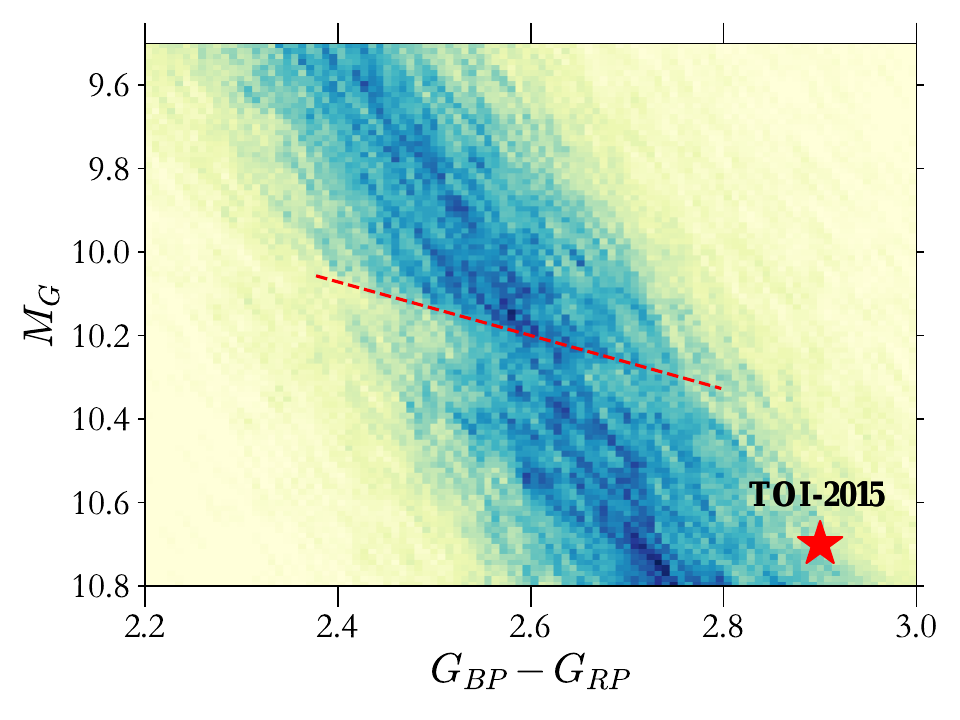}
\caption{Kernel density estimate for color-magnitude diagram based on Gaia DR3 showing M-dwarfs closer than 50 pc. The dashed line shows the upper-edge of the \textit{Jao gap} \citep{jao_gap_2018}, which is the transition zone between partially and fully-convective M-dwarfs. TOI-2015 (red star) is below this transition zone, suggesting that it is a fully-convective M-dwarf.}
\label{fig:cmd}
\end{figure}

We also place TOI-2015 in context of the transition between partially and fully-convective M-dwarf host stars \citep{limber_structure_1958, kumar_structure_1963}, by utilizing the eponymous \textit{Jao gap} to distinguish between the two regions in a color-magnitude diagram \citep[CMD;][]{jao_gap_2018, baraffe_closer_2018, feiden_gaia_2021}. We obtain a sample of stars with parallax greater than 20 mas (i.e., closer than 50 pc) queried from Gaia DR3 \citep{gaia_collaboration_gaia_2023} and subsequently cross-matched with 2MASS \citep{cutri_2mass_2003}. Finally we also apply a luminosity cut to exclude red-giants based on absolute $K_s$ magnitudes from \cite{cifuentes_carmenes_2020} as $4.7<M_{K_s}<10.1$. Figure \ref{fig:cmd} shows the resulting CMD, which shows that TOI-2015 is below the transition feature in the CMD and hence a fully-convective M-dwarf.

\subsection{Stellar Rotation and Age}
The TESS data show clear periodic out-of-transit rotational modulations. As further discussed in Section \ref{sec:transit_rv_fit}, we model the out-of-transit variability using a quasi-periodic Gaussian Process (GP) kernel following \cite{stefansson2020b}. This yields a best-fit GP periodicity of $P_{\mathrm{rot,TESS}}=6.2^{+1.0}_{-0.7} \unit{days}$. 

To gain further insight into the stellar rotation period, we also examined the available ground-based photometry from the Zwicky Transient Facility \citep[ZTF;][]{masci2019}. Figure~\ref{fig:ztf} shows the photometry in the $z_g$, $z_r$ and $z_i$ filters, along with Lomb-Scargle periodograms of the respective filters. From these plots, we see that all filters show a peak at $P_{\mathrm{rot,ZTF}} = 8.7~\unit{days}$. For both the $z_g$ and the $z_r$ filters, the peak has a false alarm probability (FAP) less than 1\%, whereas $z_i$ has a higher FAP. We attribute the higher $z_i$ FAP to the shorter time baseline and fewer data points available for the filter. Figure~\ref{fig:ztf} also shows periodograms as calculated using the TESS Systematics-Insensitive Periodogram package \citep[TESS-SIP;][]{hedges2020} for both available TESS Sectors. Although we examine Sectors 24 and 51 separately in this analysis, TESS-SIP is capable of creating periodograms while simultaneously accounting for TESS systematics within and across sectors. To build a periodogram, TESS-SIP simultaneously fits systematic regressors and sinusoidal components to the time-series data, including regularization terms to avoid overfitting. From Figure~\ref{fig:ztf}, we see that TESS Sector~24 shows a peak consistent with 8.7~days, whereas TESS Sector~51 shows a peak at 6.3~days. While running additional tests on Sector~51, we saw that the periodogram for this sector is sensitive to the maximum period chosen. We attribute this sensitivity to the lack of data available in Sector~51 due to large gaps in the TESS photometry. 

Given that the ZTF data consistently show peaks at 8.7~days in all three filters over many years, which TESS Sector~24 agrees with, we adopt a final stellar rotation period of $P_{\mathrm{rot}} = 8.7 \pm 0.9 \unit{days}$, where we have conservatively adopted a 10\% uncertainty on the rotation period.

With detections of both the photometric rotation period and the rotational velocity, we can estimate the stellar inclination. To estimate accurate posteriors for the stellar inclination $i_\star$, we use the formalism of \cite{masuda2020}, which accurately accounts for the correlated dependence between $v~\sin~i_\star$ and the equatorial velocity $v_{\mathrm{eq}}$ as calculated from $R_\star$ and $P_{\mathrm{rot}}$. As the $v~\sin~~i_\star$ measurement does not distinguish solutions between $i_\star$ and $180^\circ-i_\star$, we calculate two independent solutions between $0-90^\circ $ and $90-180^\circ$. Using our values for $P_{\mathrm{rot}}$, $v~\sin~i_\star$ and $R_\star$, we obtain two mirrored posteriors around the highest likelihood inclination of $90^\circ$, which together yield an inclination constraint of $i_\star=$\resincl. Alternatively, taking a look at only the the $0-90^\circ$ solution, our posterior constraint on the stellar inclination is $i_\star > 80^\circ$ at 68\% confidence.

Reliable age determinations for M stars is difficult due to the slow evolution of their fundamental properties \citep[e.g.,][]{laughlin1997}, and the apparent rapid transition from fast to slow rotation states \citep{newton2017}. \cite{west2015} noted that all M1V-M4V stars in their MEarth sample rotating faster than 26 days are magnetically active.
Using the gyrochronological relationship in \cite{engle2023} calibrated for M4-M6.5 stars, we infer an age of $1.1 \pm 0.1 \unit{Gyr}$. This age estimate is in line with expectations of a younger and active star, agreeing with the Ca II infrared triplet (Ca II IRT) emission seen in the HPF spectra, the flares seen in TESS, and the moderate $v\sin i = 3.2 \pm 0.6 \unit{km/s}$ value, although the age uncertainties are likely underestimated.

We note that the SED fit we ran for TOI-2015 provides an age estimate, as well, with a broad constraint of $6.8^{+4.7}_{-4.6}\unit{Gyr}$. For this study, we adopt the rotation-based age of $1.1 \pm 0.1 \unit{Gyr}$ as the primary age estimate for the system.

Additionally, we calculated the galactic space velocities $U$, $V$, and $W$ of TOI-2015 (see Table \ref{tab:stellarparam}) using the \texttt{GALPY} package \citep{bovy2015}. Following \cite{carrillo2020}, we calculate membership probabilities of 99\%, 1\%, and $\ll$1\% for TOI-2015 as a member of the galactic thin-disk, thick-disk, and halo populations, respectively. Further, from the galactic velocities, we used the BANYAN $\Sigma$ tool \citep{gagne2018} to see if TOI-2015 is a member of any known young stellar associations, from which we rule out membership to 27 well-characterized young associations within $150\unit{pc}$ with 99.9\% confidence.

\begin{figure}[ht!]
\centering
\includegraphics[width=\columnwidth]{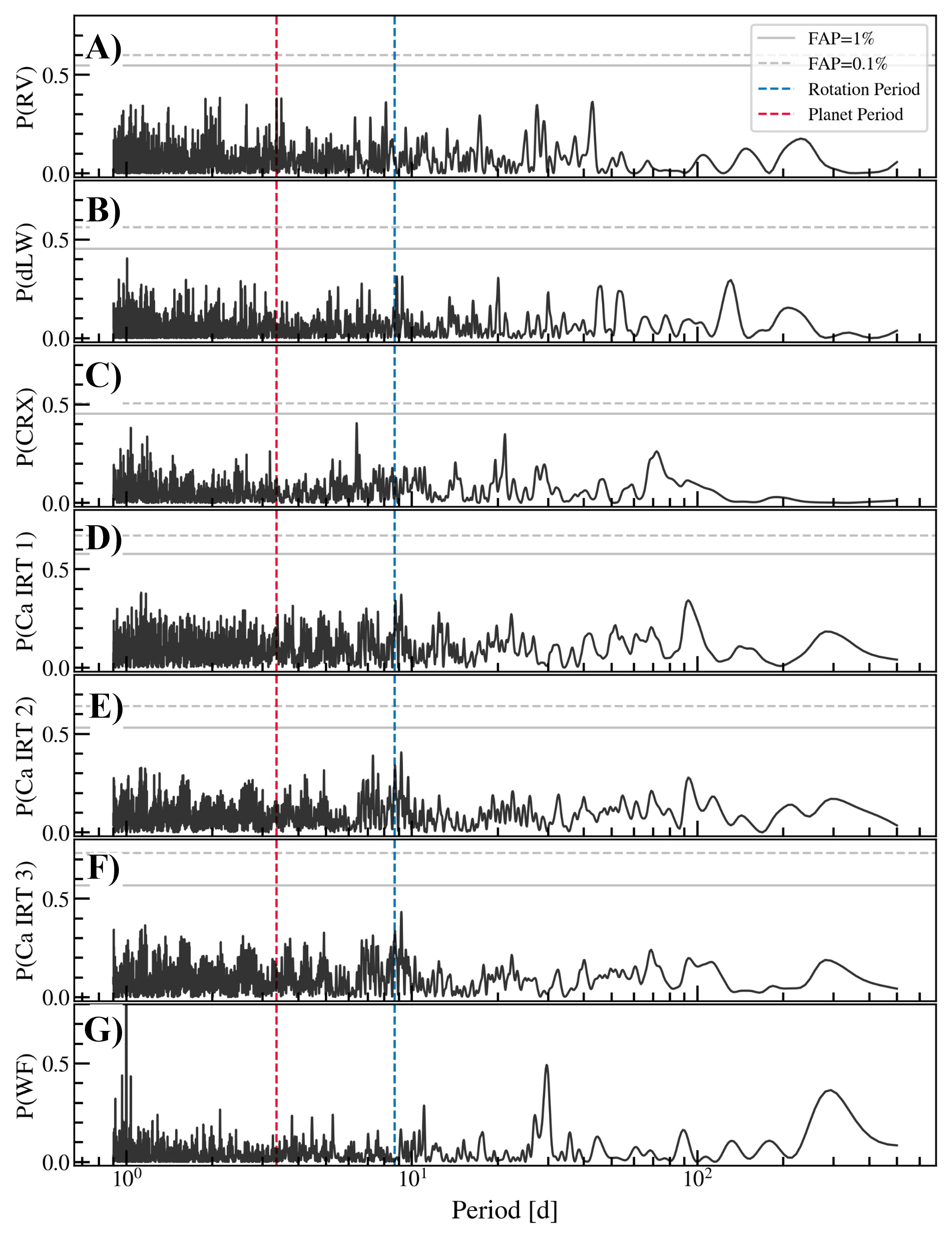}
\caption{Generalized Lomb Scargle periodograms of the HPF RVs and HPF activity indicators. \textit{A}) HPF RVs, \textit{B}) HPF differential line width (dLW) indicator, \textit{C}) chromatic activity index (CRX), \textit{D-F}) Ca II infrared triplet line indicators, \textit{G}) window function (WF) of the observations. The known planet period and the stellar rotation period are denoted with the vertical blue and red dashed lines, respectively. False alarm probabilities of 1\% and 0.1\% calculated using a bootstrap method are denoted with the grey solid and grey dashed lines, respectively. Although we see a hint of a peak near the known planet period, it is not significant (FAP>1\%). For the activity indicators, we see no significant peaks at the known planet or stellar rotation periods. Periodograms A-F are normalized using the formalism in \cite{zechmeister2009}, while the window function in panel G is normalized such that the highest peak has a power of 1.}
\label{fig:periodograms}
\end{figure}

\subsection{Spectroscopic Stellar Activity Indicators}
Magnetic activity can create planet-like signals in RV data \citep[e.g.,][]{robertson2014}. To probe corresponding periodicities in the HPF activity indicators, Figure~\ref{fig:periodograms} compares Generalized Lomb Scargle periodograms of the HPF RVs along with the HPF differential line width (dLW) indicator, chromatic index (CRX), and the Ca II near-infrared triplet indicators (Ca II IRT)\footnote{The wavelength region definitions for our Ca II IRT activity index are listed in \cite{stefansson2020b}.}. To generate the periodograms, we used the \texttt{astropy.timeseries} package, which we also used to calculate associated false-alarm probabilities using the \texttt{bootstrap} method available in the package. The TOI-2015 periodograms show small hints of a peak at the known planet period, but it is not considered significant with FAP>1\%.

\begin{figure*}[t!]
\centering
\includegraphics[width=0.8\textwidth]{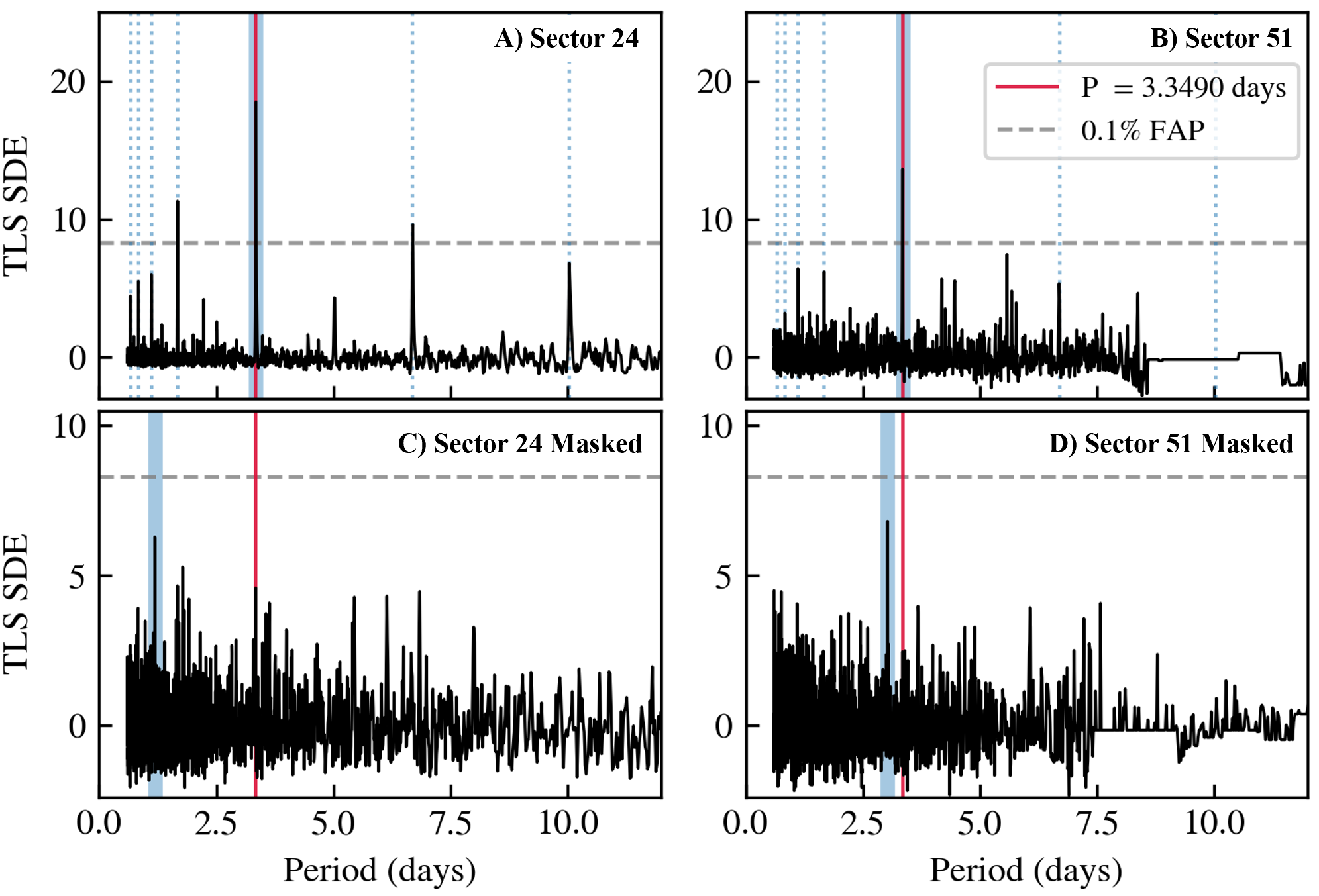}
\caption{Transit Least Squares Signal Detection Efficiency (TLS SDE) power spectra as a function of orbital period for TESS Sectors 24 and 51. The solid blue bands indicate the maximum peak of each power spectrum. The blue dotted lines in panels \textit{A} and \textit{B} are the integer harmonics of the 3.3490 day orbital period of TOI-2015b in red. The grey dashed lines represent 0.1\% false alarm probability (FAP). We do not detect signals of an additional transiting companion in the TESS data.}
\label{fig:tls}
\end{figure*}

From visual inspection of the HPF spectra, we see clear evidence of chromospheric emission in the cores of the Ca II IRT lines, confirming that TOI-2015 is an active star. However, this chromospheric variability does not show an apparent corresponding periodicity with either the known planet period or the stellar rotation period, as seen in Figure~\ref{fig:periodograms}. The lack of significant peaks in either the RV or dLW periodograms at $P_\mathrm{rot} = 8.7~\unit{days}$ suggests that the RV impact of stellar activity is not a dominant signal for this star given our median RV precision level of $\sim15\unit{m/s}$. Additionally, we show the window function of the RVs in Figure~\ref{fig:periodograms}, which reveals a peak at the lunar cycle of $P\sim28~\unit{days}$, as the HPF observations were preferentially executed during bright time in the HET queue. Table \ref{tab:rvs} in the Appendix lists the values of the RVs and the activity indicators used in this work.

\newpage

\section{Transit, RV, and TTV Modeling}
\label{sec:planetparams}

\subsection{Search for Additional Transiting Planets in TESS}
\label{sec:transit_search}
To identify potential additional transiting planets, we used the \texttt{transitleastsquares} \citep[\texttt{TLS};][]{hippke2019} code to look for periodic transit-like features in the TESS photometry\footnote{\url{https://transitleastsquares.readthedocs.io/en/latest/index.html}}. We extracted the PDCSAP 2 minute cadence data using the \texttt{lightkurve} package \citep{lightkurve}. Using \texttt{TLS}, we removed outliers and detrended the photometry with a Savitzky-Golay high-pass filter of window length 501~cadences. We chose this $\sim~$17~hour window to help remove out-of-transit variability, while having a minimal impact on the transit features. Due to the 2 year gap between Sectors 24 and 51, we chose to analyze each sector individually.

\begin{figure*}[htp!]
\centering
\vspace{.3cm}
\includegraphics[width=0.9\textwidth]{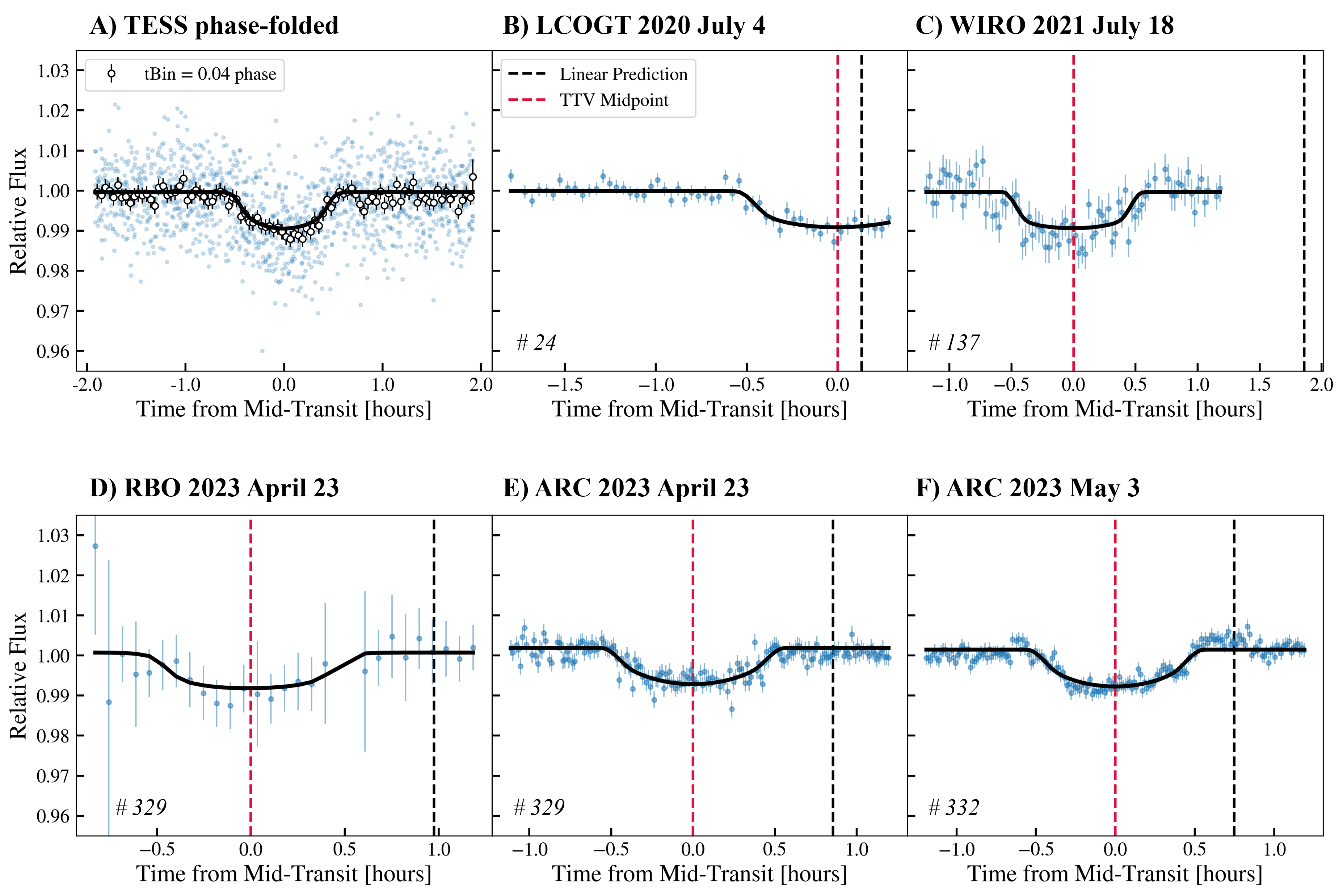}
\caption{TOI-2015 phased photometry from \texttt{juliet} joint TTV fit. Data points are plotted in blue. The best-fit transit models are plotted with solid black lines. The vertical black lines represent the expected linear midpoint predictions, and the vertical red lines are the true TTV midpoints. The transit number (first TESS transit is \#0) is labeled in the bottom left corner of each ground-based transit. \textit{A}) Phase-folded photometry of 12 TESS transits. White points are NBin~=~15 binned data. \textit{B}) LCOGT partial transit photometry. \textit{C}) WIRO full transit photometry. \textit{D}) RBO full transit photometry. \textit{E}) ARC 2023 April full transit photometry. \textit{F}) ARC 2023 May full transit photometry.}
\label{fig:phot_overview}
\end{figure*}

Figure~\ref{fig:tls} shows the \texttt{TLS} power spectra for TESS Sectors 24 and 51 before and after masking out the TOI-2015b transits. Using \texttt{TLS}, we find the 3.3490 day period of planet b is featured strongly in the resulting power spectra (above the 0.1\% FAP line). To identify other potential transiting companions, we masked out all TOI-2015b in-transit regions in the TESS light curve, with windows 3 times wider than the transit duration to account for TTVs. After searching the masked data, the TLS power spectrum revealed no significant peaks (FAP < 0.1\%, correlating to a \texttt{TLS} Signal Detection Efficiency (SDE) > 8.3) indicative of additional transiting planets.

From this analysis, we conclude there is evidence for only one transiting planet in the system, and that the orbiting companion causing the TTVs that we observe is not transiting in the available TESS data.

\begin{figure*}[htp!]
\centering
\vspace{.3cm}
\includegraphics[width=0.97\textwidth]{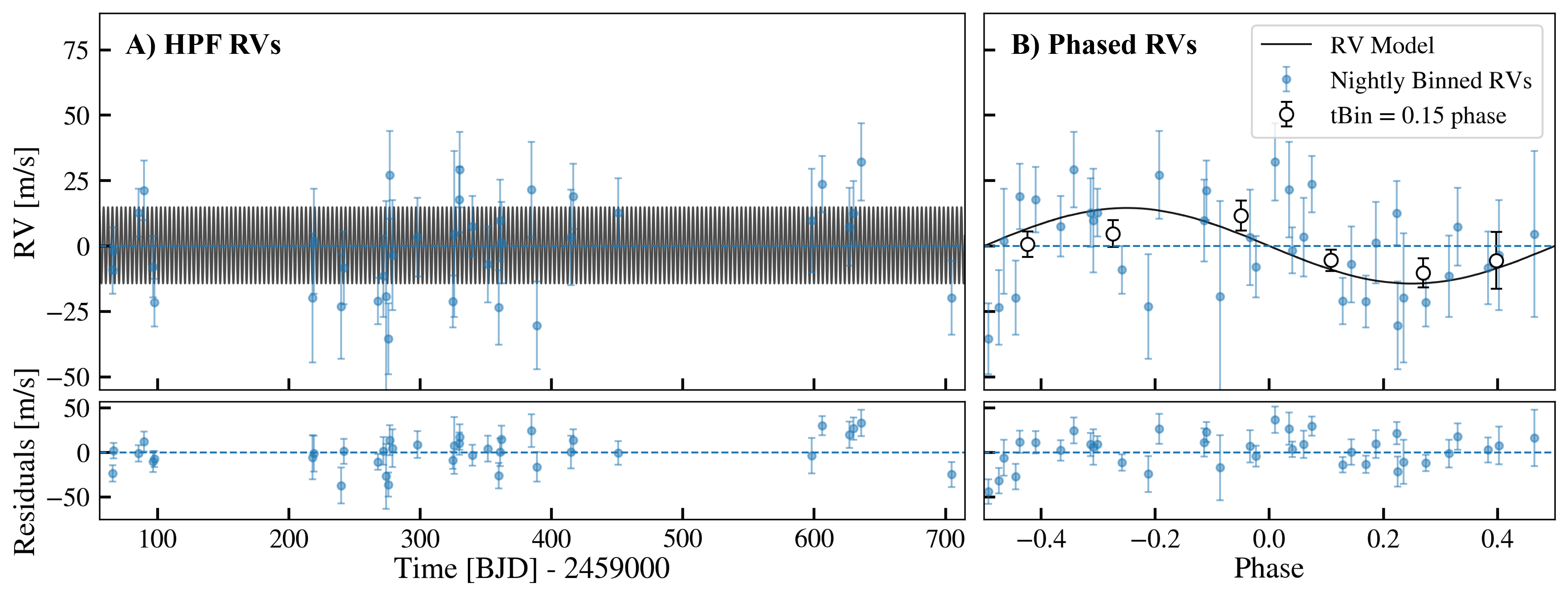}
\caption{HPF radial velocity measurements. RVs (blue points) are plotted with respect to the baseline (horizontal dashed line) and the best fit model (bold black line). \textit{A}) RVs as a function of time. \textit{B}) RVs phase folded at the period of TOI-2015b. Residuals are displayed in the corresponding lower panels. The RV data are listed in Table~\ref{tab:rvs} in the Appendix. TOI-2015b is evident in the RVs, but there are no additional planetary signals.}
\vspace{.3cm}
\label{fig:rv}
\end{figure*}

\subsection{Joint Transit and RV Fit}
\label{sec:transit_rv_fit}
As such, we jointly modeled the TESS and ground-based photometry and HPF radial velocities using the \texttt{juliet} software package \citep{espinoza2018b} assuming a single transiting planet. We utilize dynamic nested sampling with the \texttt{dynesty} sampler \citep{speagle2019} to estimate the posterior parameter space of our joint fit. We parameterized the transits using the $r_1$ and $r_2$ parameters from \cite{espinoza2018b}, which probe only the full range of physically possible values for scaled radius $R_{p}/R_{*}$ and impact parameter $b$. We also used stellar
density $\rho_*$ and the \cite{kipping2013} quadratic limb darkening parameterizations $q_1$ and $q_2$ as parameters in the fit.

\newpage
\clearpage
Because the TESS photometry showed evidence of periodic out-of-transit variability, we incorporate Gaussian Processes (GP) in our model. To fit for the periodic out-of-transit variability, we applied the \texttt{celerite} \citep{Foreman-Mackey2017} quasi-periodic kernel to the TESS photometry. GPs were not needed for ground-based data, because these few-hour observations are not long enough for a significant rotational modulation signal to be detected. The quasi-periodic kernel in \texttt{celerite} is parameterized by the following hyperparameters: a GP amplitude \emph{B}, an additive factor impacting the amplitude \emph{C}, the length-scale of the exponential part of the GP \emph{L}, and the GP period $P_{GP}$. To account for transit timing variations, we used \texttt{juliet}'s TTV modeling feature. We used a uniform prior with a 0.2 day window around the expected transit midpoint for each of the 17 transits from the TESS and ground-based data. 

Figures \ref{fig:phot_overview} and \ref{fig:rv} summarize the results from the fit, showing the best-fit joint model for the photometry, and RVs, respectively. Additionally, Figure~\ref{fig:tess_transits} in the Appendix shows the individual TESS transits along with the best fit transit models accounting for the TTVs. The priors and final parameters are summarized in Table \ref{tab:julietfitparams}, and Table \ref{tab:ttvs} in the Appendix shows the individually derived TTV midpoints for each transit. Those TTV midpoints are further fit and discussed in Section \ref{sec:TTV} to constrain the possible parameters of the second planet candidate.

The fit above assumed a circular orbit for planet b. In addition to the circular fit, we experimented with running a fit where we let the eccentricity vary. In doing so, we obtain a best-fit eccentricity of $e~=~0.23~\pm~0.07$ with a log evidence value of $\ln(Z)~=~88489.91$. Whereas the circular fit has a log-evidence value of $\ln(Z)~=~88488.80$. As this log evidence value is only $\Delta \ln (Z) = 1.1$ or $\Delta Z = 3.0$ larger, we consider the models to be statistically indistinguishable. Here we have followed the suggestion in \cite{espinoza2019} of requiring at least $\Delta \ln (Z) = 2$ or equivalently $\Delta Z = 7.4$ as a threshold for statistical significance. As such, we elect to adopt the posteriors from the simpler circular fit, as listed in Table \ref{tab:julietfitparams}.

\begin{deluxetable*}{llcc}
\tabletypesize{\scriptsize}
\vspace{-0.3cm}
\tablecaption{Median values and 68\% credible intervals from \texttt{juliet} joint photometry and radial velocity fit. TTV midpoints are displayed in Table \ref{tab:ttvs} in the Appendix. Prior denotations: $\mathcal{N}(\mu,\sigma)$ denotes a normal prior with mean $\mu$, and standard deviation $\sigma$; $\mathcal{U}(a,b)$ denotes a uniform prior with a start value $a$ and end value $b$, and $\mathcal{J}(a,b)$ denotes a Jeffreys prior truncated between a start value $a$ and end value $b$.}
\label{tab:julietfitparams}
\tablehead{
\colhead{Parameter}                                                             &  \colhead{Description}                             &  \colhead{Prior}              &   \colhead{TOI-2015b}}
\startdata	
\hline
\multicolumn{4}{l}{\hspace{-0.3cm} \textbf{Orbital Parameters:}} \\
$P_{\mathrm{orb}}$                                                                             &  Orbital Period (days)                             & $\mathcal{U}(3.33,3.36)$       & $3.348968_{-0.000004}^{+0.000004}$ \\
$T_C$                                                                           &  Transit Midpoint - 2458000 $(\mathrm{BJD_{TDB}})$ & $\mathcal{U}(955.9,956.1)$     & $956.0226_{-0.0009}^{+0.0009}$     \\
$r_1$                                                                           &  \cite{espinoza2018b} Parameterization for Impact Parameter b                 & $\mathcal{U}(0,1)$             & $0.797_{-0.022}^{+0.020}$          \\
$r_2$                                                                           &  \cite{espinoza2018b} Parameterization for Scaled Radius $R_{p}/R_{*}$                 & $\mathcal{U}(0,1)$             & $0.0933_{-0.0020}^{+0.0020}$       \\
$R_{p}/R_{*}$                                                                   &  Scaled Radius                                     & .....                          & $0.0933_{-0.0020}^{+0.0020}$       \\
$a/R_*$                                                                         &  Scaled Semi-major axis                            & .....                          & $19.46_{-0.51}^{+0.47}$            \\
$b$                                                                             &  Impact Parameter                                  & .....                          & $0.696_{-0.033}^{+0.031}$          \\
$e$                                                                             &  Eccentricity                                      & 0                              & 0                                  \\
$\omega$                                                                        &  Argument of Periastron (\textdegree)              & 90                             & 90                                 \\
$K$                                                                             &  RV semi-amplitude ($\mathrm{m/s}$)                  & $\mathcal{U}(0,100)$           & $14.4_{-3.6}^{+3.5}$               \\
\multicolumn{4}{l}{\hspace{-0.3cm} \textbf{Instrumental Terms:}}                \\
$q_1$                                                                           &  Limb-darkening parameter                          & $\mathcal{U}(0,1)$             & $0.70_{-0.23}^{+0.20}$             \\
$q_2$                                                                           &  Limb-darkening parameter                          & $\mathcal{U}(0,1)$             & $0.60_{-0.29}^{+0.25}$             \\
$\sigma_{\mathrm{phot}}$                                                        &  Photometric jitter ($\unit{ppm}$)                 & $\mathcal{J}(1,5000)$          & $8.9_{-6.8}^{+37.3}$               \\
$\mu_{\mathrm{phot}}$                                                           &  Photometric baseline                              & $\mathcal{N}(0,0.1)$           & $0.0003_{-0.0011}^{+0.0011}$       \\
$\sigma_{\mathrm{HPF}}$                                                         &  HPF RV jitter (m/s)                               & $\mathcal{J}(0.1,100)$         & $4.75_{-4.16}^{+4.86}$             \\
$\gamma$                                                                        &  HPF RV offset (m/s)                               & $\mathcal{U}(-60,60)$          & $-2.2_{-2.3}^{+2.3}$               \\
\multicolumn{4}{l}{\hspace{-0.3cm} \textbf{TESS Quasi-Periodic GP Parameters:}} \\
$B$                                                                             &  GP Amplitude ($\unit{ppm^2}$)                     & $\mathcal{J}(10^{-6},1)$       & $1.7\mathrm{e}{-5}_{-0.000004}^{+0.000006}$  \\
$C$                                                                             &  GP Additive Factor                                & $\mathcal{J}(10^{-3},10^{3})$  & $0.028_{-0.025}^{+0.25}$           \\
$L$                                                                             &  GP Length Scale (days)                            & $\mathcal{J}(10^{-2}, 10^{5})$ & $3.5_{-1.0}^{+1.5}$                \\
\multicolumn{4}{l}{\hspace{-0.3cm} \textbf{Additional Derived Parameters:}}     \\
$R_{p,\oplus}$                                                                         &  Planet radius ($R_\oplus$)                       & .....                          & $3.39_{-0.13}^{+0.14}$             \\
$R_{p,\mathrm{J}}$                                                                         &  Planet radius ($R_{\mathrm{J}}$)                     & .....                          & $0.302_{-0.012}^{+0.012}$          \\
$\delta_{p, \mathrm{K2}}$                                                       &  Transit depth                                     & .....                          & $0.0087_{-0.00037}^{+0.00038}$     \\
$a$                                                                             &  Semi-major axis$^a$ (AU)           & .....                          & $0.0301_{-0.0012}^{+0.0013}$       \\
$i$                                                                 &  Transit inclination $(^{\circ})$                               & .....                          & $87.95_{-0.14}^{+0.13}$            \\
$T_{\mathrm{eq},0.0}$                                                           &  Equilibrium temp. $a=0.0$ (K)                       & .....                          & $512.0_{-11.0}^{+11.0}$            \\
$T_{\mathrm{eq},0.3}$                                                           &  Equilibrium temp. $a=0.3$ (K)                       & .....                          & $358.5_{-7.7}^{+7.8}$              \\
$S$                                                              &  Insolation Flux ($S_{\oplus}$)                                   & .....                          & $11.47_{-0.95}^{+1.0}$             \\
$T_{14}$                                                                 &  Transit duration                                  & .....                          & $0.0463_{-0.0010}^{+0.0010}$ (days)       \\
$T_{23}$                                                                 &  Transit duration (days)                                  & .....                          & $0.0318_{-0.0018}^{+0.0018}$       \\
$\tau$                                                                   &  Ingress/egress duration (days)                           & .....                          & $0.00719_{-0.00056}^{+0.00065}$    \\
$m_p$                                                              &  Planet mass ($M_\oplus$)                                       & .....                          & $16.4_{-4.1}^{+4.1}$               \\
$\rho_p$                                                             &  Planet density$^b$ ($\mathrm{g/cm^3}$)                                      & .....                          & $2.32_{-0.37}^{+0.38}$               \\
\enddata
\tablenotetext{a}{Calculated from $a/R_*$ and $R_*$}
\tablenotetext{b}{Calculated from $m_p$ and $r_p$}
\end{deluxetable*}

\subsection{Joint TTV and RV Fit}
\label{sec:TTV}

\begin{deluxetable*}{llcccc}
\tabletypesize{\scriptsize}
\vspace{-0.3cm}
\tablecaption{Median values and 68\% credible intervals from \texttt{jnkepler} joint TTV and RV fit.}
\tablehead{
\colhead{Parameter}&  \colhead{Description}          &   \colhead{Prior}        &  \colhead{2:1 Solution} &   \colhead{3:2 Solution} &   \colhead{4:3 Solution} }
\startdata	
\hline
\multicolumn{6}{l}{\hspace{-0.3cm} \textbf{Parameters of Planet b:}} \\
$P_b$         & Orbital Period (days)         &  $\mathcal{U}(3.149, 3.549)$ & $3.3452_{-0.0004}^{+0.0004}$ & $3.3454$  &  $3.3448$ \\
$t_{0,b}$     & Transit Midpoint - 2458000 $(\mathrm{BJD_{TDB}})$  & $\mathcal{U}(955.934, 956.134)$ & $956.0342_{-0.0017}^{+0.0017}$   & $956.0326$  &  $956.0329$ \\
$e_b\cos\omega_b$    & Eccentricity and Argument of Periastron &   $e_b\sim \mathcal{N}_t(0, 0.1, 0)$    & $0.077_{-0.097}^{+0.083}$ & $0.036$  & $0.008$     \\
$e_b\sin\omega_b$    & Eccentricity and Argument of Periastron &   $\omega_b \sim \mathcal{U}_w(-\pi, \pi)$ & $0.035_{-0.126}^{+0.095}$ & $0.027$ & $0.001$\\
$M_b$         & Mass ($M_\oplus$)   & $\mathcal{U}(0,100)$                           & $13.3_{-4.5}^{+4.7}$   &    $11.2$   &  $4.8$ \\
\multicolumn{6}{l}{\hspace{-0.3cm} \textbf{Parameters of Candidate Planet c:}} \\
$P_c$         & Orbital Period (days)   & $\mathcal{U}(\hat{P}_c-0.2, \hat{P}_c+0.2)$\tablenotemark{$_*$} &  $6.7179_{-0.0073}^{+0.0107}$\tablenotemark{$_\dagger$}  & $5.0524$ &  $4.4842$ \\
$\tau_{0,c}=(t_{0,c}-2458955)/P_c$     
& Transit Midpoint in Units of $P_c$  & $\mathcal{U}_w(0,1)$ & $0.33_{-0.16}^{+0.15}$\tablenotemark{$_\dagger$}  & $0.21$  &  $0.09$ \\
$e_c\cos\omega_c$    & Eccentricity and Argument of Periastron &   $e_c\sim \mathcal{N}_t(0, 0.1, 0)$    & $0.003_{-0.065}^{+0.078}$ & $0.05$ & $0.03$    \\
$e_c\sin\omega_c$    & Eccentricity and Argument of Periastron &   $\omega_c \sim \mathcal{U}_w(-\pi, \pi)$ & $-0.035_{-0.049}^{+0.048}$ & $0.00$ & $-0.01$ \\
$M_c$         & Mass ($M_\oplus$)    & $\mathcal{U}(0,100)$                          & $6.8_{-2.3}^{+3.5}$   &    $2.4$    &  $1.3$ \\
\multicolumn{6}{l}{\hspace{-0.3cm} \textbf{Stellar Parameters:}}     \\
$M_\star$ & Mass ($M_\odot$) & $\mathcal{N}_t(0.34, 0.025, 0.25)$ & $0.340_{-0.024}^{+0.025}$ & $0.34$ & $0.34$\\
\multicolumn{6}{l}{\hspace{-0.3cm} \textbf{RV Model Parameters:}}          \\
$\rho$        & Undamped Period of the Oscillator (days)  & $\mathcal{N}_t(8.7, 0.87, 0.7)$ &  $8.7_{-0.8}^{+0.9}$  & $8.8$ & $8.9$ \\
$\ln\tau$     & Log of Damping Timescale of the Process (days)   & $\mathcal{U}(-3,6)$  & $0.9_{-2.9}^{+3.6}$  & $1.7$ & $-0.83$\\
$\ln\sigma$   & Log of Standard Deviation of the Process (m/s)   & $\mathcal{U}(-3,5)$  & $1.1_{-2.7}^{+1.2}$  & $1.4$ & $1.2$\\
$\gamma$           & HPF RV Offset (m/s)   & $\mathcal{U}(-30,30)$            & $2.3_{-2.9}^{+3.1}$  & $3.3$ & $3.1$\\
$\ln\sigma_{\rm jit}$ & Log of HPF RV Jitter (m/s) &  $\mathcal{U}(-1,5)$     & $1.7_{-1.6}^{+0.7}$  & $1.7$ & $2.2$
\enddata
\tablecomments{Orbital elements are the Jacobi elements defined at $\mathrm{BJD}-2458000=955$. For the 3:2 and 4:3 solutions, only the median values are shown because MCMC chains showed insufficient convergence.
Prior denotations: $\mathcal{N}(\mu,\sigma)$ denotes a normal prior with mean $\mu$, and standard deviation $\sigma$; $\mathcal{N}_t(\mu,\sigma,l)$ denotes $\mathcal{N}(\mu,\sigma)$ truncated at the lower bound $l$; $\mathcal{U}(a,b)$ denotes a uniform prior with a start value $a$ and end value $b$; $\mathcal{U}_w(a,b)$ denotes a ``wrapped'' uniform prior where $a$ and $b$ are treated as the same points, and $\mathcal{J}(a,b)$ denotes a log-uniform prior truncated between a start value $a$ and end value $b$.}
\tablenotemark{$_*$}{The values of $\hat{P}_c$ are $2P_b$, $(3/2)P_b$, and $(4/3)P_b$ for the 2:1, 3:2, and 4:3 solutions, respectively.}
\tablenotemark{$_\dagger$}{Parameters showing poorer convergence.}
\label{tab:jointttvrv}
\end{deluxetable*}

Next we searched for two-planet solutions consistent with the TTVs and RVs of TOI-2015b. Large sinusoidal TTVs are typical in systems near mean-motion resonances, and so here we focus on solutions for which the period ratio of the second planet candidate and TOI-2015b is around 2:1, 3:2, and 4:3. Throughout this section, the TTVs are computed with full $N$-body integrations, assuming coplanar orbits and ignoring any relativistic corrections. Because the 2 minute cadence TESS photometry shows no evidence of a second transiting planet (including at the given period ratios; see Section \ref{sec:transit_search}), it is unclear whether the assumption of coplanarity is correct. However, the lack of transit provides little information in this aspect; given TOI-2015b's impact parameter of $\approx 0.7$, an outer planet candidate with a period ratio $\gtrsim 1.7$ does not transit even assuming perfect coplanarity. It is not our goal here (nor is it practical) to explore all possible period ratios and mutual orbital inclinations for the second planet candidate. Rather, we would like to know whether reasonable two-planet solutions exist---and if so, what some of them look like---to guide future observations to better characterize this system, as well as to check the robustness of the RV mass that was derived ignoring the second planet candidate.

First, we combined {\tt TTVFast} \citep{deck2014} and {\tt MultiNest} \citep{2009MNRAS.398.1601F, 2014A&A...564A.125B}, as described in \citet{2017AJ....154...64M}, to find optimal solutions around 2:1, 3:2, and 4:3 resonances, setting narrow priors on the period of the outer planet candidate. The prior ranges were chosen so that the expected super period \citep{2012ApJ...761..122L} matches the $\sim430\,\mathrm{days}$ periodicity seen in the data. We obtained the maximum likelihood parameters from the {\tt MultiNest} runs, and performed chi-squared optimization starting from the solution, using the Trust Region Reflective method implemented in {\tt scipy.optimize.curve\_fit}. In all cases, we found acceptable solutions that provide $\chi^2 \approx 10$ for 17 data points and 10 free parameters (i.e., $\chi^2/\mathrm{dof}=10/7\approx1.4$).

We then perform a joint TTV-RV fit for each of the three period ratios. Here we obtain samples from the joint posterior distribution for the system parameters $\theta$ conditioned on the TTV data $t^{\rm obs}$ and RV data $v^{\rm obs}$:
\begin{equation}
p(\theta | t^{\rm obs}, v^{\rm obs}) \propto \mathcal{L}(\theta) \, \pi (\theta).
\end{equation}
Here $\pi$ denotes the prior probability distribution for $\theta$ that is assumed to be separable for each parameter (see Table~\ref{tab:jointttvrv}). We adopt the following log-likelihood function: 
\begin{align}
\notag
\ln\mathcal{L}(\theta) &= 
-{\frac{1}{2}} \sum_i\left(\frac{t^{\rm obs}_i - t^{\rm model}_i}{\sigma^{\rm obs}_i}\right)^2 \\
\notag
    &-{\frac{1}{2}}\sum_{i,j} (v^{\rm obs}_i-v^{\rm model}_i) (K^{-1})_{ij}  (v^{\rm obs}_j-v^{\rm model}_j) \\
    &-{\frac{1}{2}}\ln\mathrm{det}\,K + \mathrm{constant},
\end{align}
where $t^{\rm model}$ is the modeled transit midpoint including the TTV and $v^{\rm model}$ is the expected radial velocity, both computed via full $N$-body integration. The $N$-body code is implemented in {\tt JAX} \citep{jax2018github} to enable automatic differentiation with respect to the input orbital elements and mass ratios \citep[see also][]{agol2021}, and is available through GitHub as a part of the {\tt jnkepler} package (Masuda 2023, in prep.).\footnote{\url{https://github.com/kemasuda/jnkepler}} 
In our model, we assume that the TTV errors follow independent zero-mean Gaussian distributions whose widths are the assigned timing errors as listed in Table~\ref{tab:ttvs}. In evaluating the likelihood for the RV data, we use a Gaussian process to model a correlated noise component that may exist given the activity of the star. Here we adopt a simple harmonic oscillator covariance function $K$ that corresponds to the following power spectral density:
\begin{equation*}
S(\omega) = \sqrt{\frac{2}{\pi}} \frac{S_0\,\omega_0^4}
{(\omega^2-{\omega_0}^2)^2 + {\omega_0}^2\,\omega^2/Q^2},
\end{equation*}
which was parameterized by the undamped period of the oscillator $\rho = 2\,\pi / \omega_0$, the standard deviation of the process $\sigma = \sqrt{S_0\,\omega_0\,Q}$, and the quality factor $Q$ \citep{celerite1}. We also included a jitter term whose square was added to the diagonal elements of the covariance matrix. The Gaussian process log-likelihood was evaluated using the {\tt JAX} interface of {\tt celerite2} \citep{celerite, celerite2}. Leveraging the differentiable $\mathcal{L}(\theta)$, we sample from $p(\theta|t^{\rm obs}, v^{\rm obs})$ using Hamiltonian Monte Carlo and the No-U-Turn Sampler \citep{DUANE1987216, 2017arXiv170102434B} as implemented in {\tt NumPyro} \citep{bingham2018pyro, phan2019composable}. We run six chains in parallel, setting the target acceptance probability to be 0.95 and the maximum tree depth to be 11. 

\begin{figure*}[htb!]
\centering
\includegraphics[width=0.97\textwidth]{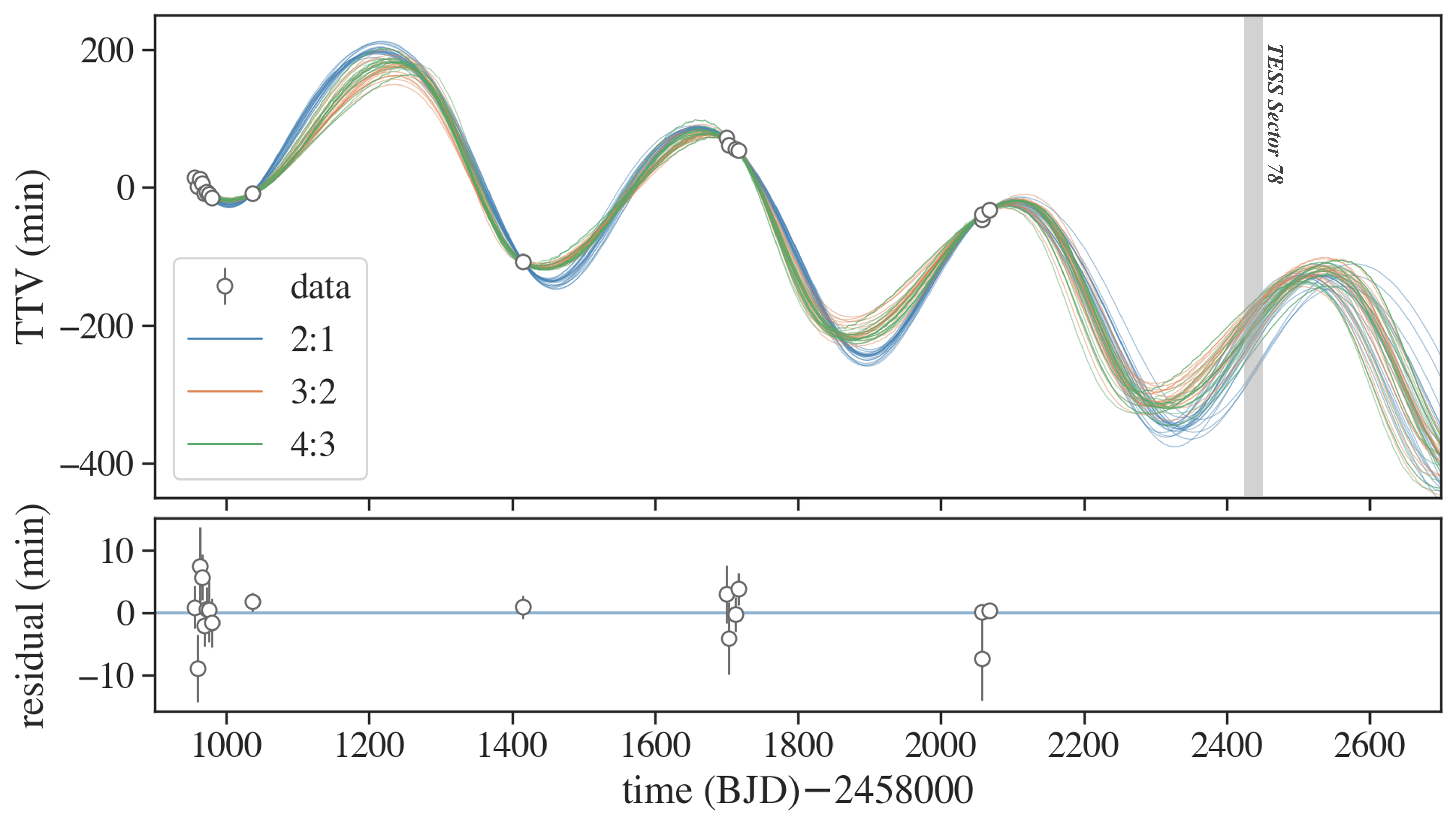}
\caption{Transit Timing Variations (TTVs) of TOI-2015b as a function of time (filled circles). 20 models corresponding to stable posterior samples from each of the 2:1, 3:2, and 4:3 joint TTV-RV solutions are shown with thin blue, orange, and green lines, respectively. The residuals in the bottom panel are shown with respect to the maximum-likelihood 2:1 solution. Here the TTVs are shown with respect to the ephemeris of $t_0(\mathrm{BJD})=2458956.0238$ and $P=3.3489393\,\mathrm{days}$ obtained by linear fitting of the observed transit times. The linear trend seen in the posterior TTV models in this figure indicates that the mean orbital period inferred from the dynamical modeling differs from the value estimated from the available transit times. TESS will observe TOI-2015 again from 2024 April 23 to 2024 May 21 in Sector 78, denoted by the gray region in the figure.}
\label{fig:ttvprediction}
\end{figure*} 

\begin{figure*}[htb!]
\centering
\includegraphics[width=0.95\textwidth]{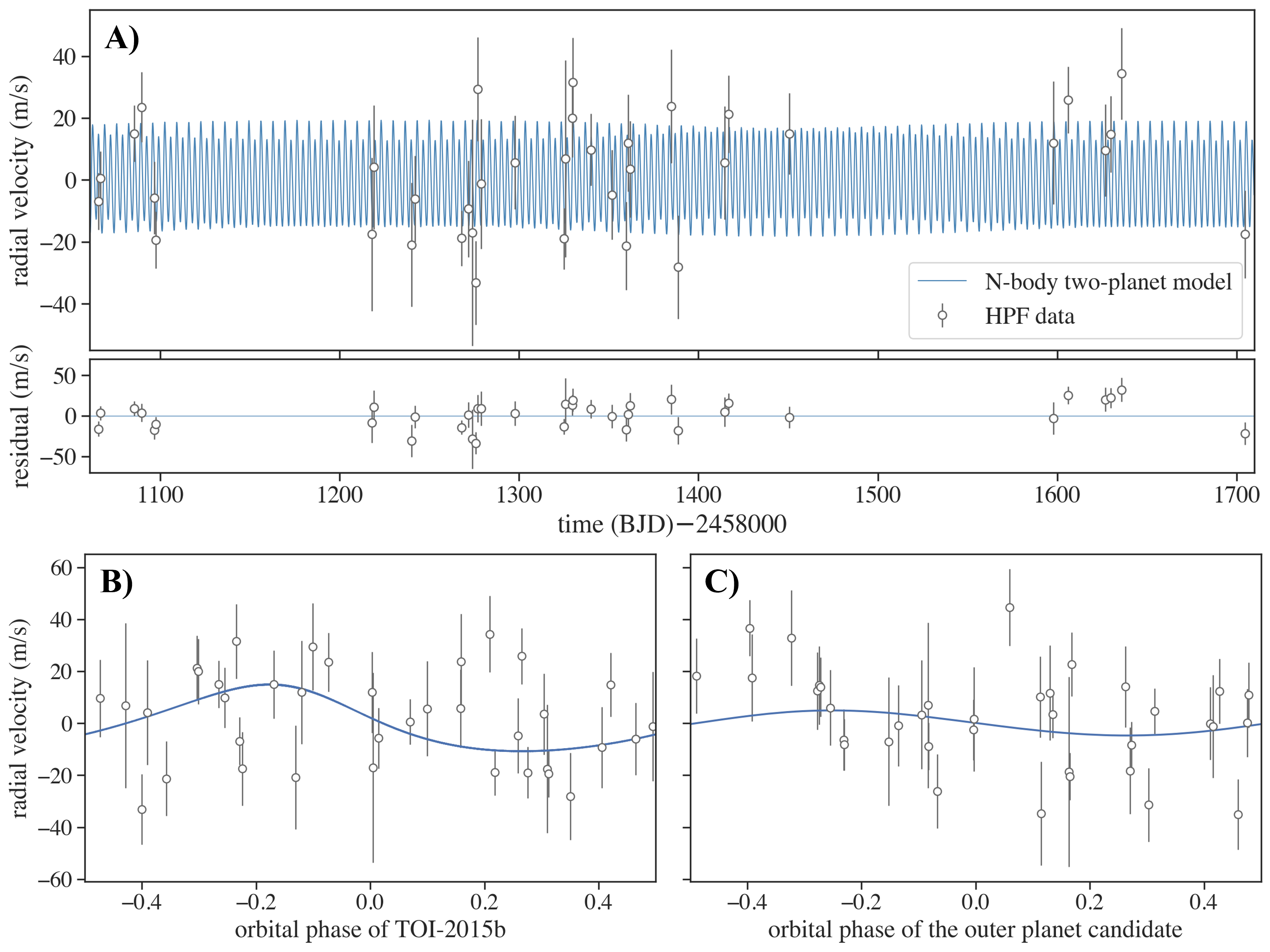}
\caption{
\textit{A)} The maximum-likelihood 2:1 solution from the joint TTV and RV fit (blue solid line) compared with the HPF data (white circles with error bars). \textit{B-C)} The contribution of each planet to the RV signal in the above solution is shown as a function of the orbital phase. Here the signal due to each planet is computed by setting the mass of the other planet to be zero in the $N$-body model. In the right panel, the RV signal due to the inner planet is subtracted from the data. Note that the stellar RVs are computed via direct $N$-body integration in our joint TTV and RV fit, rather than by adding two independent Keplerian signals; thus the RV signals do not have fixed periods in reality.
}
\label{fig:nbodyrv}
\end{figure*} 

After obtaining posterior samples, we performed longer-term integration for a subset of the samples to remove solutions that quickly become unstable. We use the {\tt mercurius} integrator \citep{reboundmercurius} in the {\tt REBOUND} package \citep{rebound} to integrate $10^8$ inner orbits (i.e., $\sim1\,\mathrm{Myr}$). We flagged a solution to be unstable if the semi-major axis of TOI-2015b changed by more than $1\%$ during the integration, in which case the period ratio of the two planets changed by more than $\sim 10\%$, making the current near-resonant architecture implausible \citep{2023AJ....165...33D}.

The results of the joint TTV-RV modeling are summarized in Table~\ref{tab:jointttvrv}. For the 2:1 solution, the effective number of samples reached 50 and the split Gelman--Rubin statistic achieved $\hat{R}<1.1$ \citep[][Chapter 11]{BB13945229} after running 20,000 steps for most parameters except for the transit midpoint ($\tau_c$) and period ($P_c$) of the second planet candidate. The poorer convergence for these two parameters is due to strong and complicated degeneracy between these parameters and eccentricity vectors (see Figure~\ref{fig:ttv_corner}), and their summary statistics may be of limited accuracy. We tested stability for 10,000 randomly chosen posterior samples and found that $>90\%$ survived. Thus the 2:1 solution qualifies as a valid explanation for the TTV and RV data.
From the joint TTV-RV fit, we inferred the masses of $m_b = 13.3^{+4.7}_{-4.5}\,M_\oplus$ for TOI-2015b and $m_c = 6.8^{+3.5}_{-2.3}\,M_\oplus$ for the outer planet candidate. This mass for TOI-2015b is consistent with the value from our \texttt{juliet} RV fitting assuming a single planet of $m_b = 16.4\pm4.1 M_\oplus$ from Table \ref{tab:julietfitparams}.
The mass constraint from the joint TTV-RV fit is slightly less precise than the \texttt{juliet} value. This is not necessarily surprising given the additional degrees of freedom (i.e., eccentricity of TOI-2015b, mass and orbital elements of the second planet candidate) introduced in the joint fit.
The summary statistics in Table~\ref{tab:jointttvrv} are based on the samples that were classified to be stable. The TTV models corresponding to 20 of those samples are shown in Figure~\ref{fig:ttvprediction} with blue solid lines. The RV model corresponding to the maximum likelihood sample is shown in Figure~\ref{fig:nbodyrv}.

In contrast to the 2:1 solution, for the 3:2 and 4:3 solutions, we could not obtain well-mixed chains. This appears to be due to strong multimodality in the ephemeris of the second planet candidate ($\tau_c$ and $P_c$). Thus we do not have well-defined estimates for the uncertainties of the parameters, and quote only medians of the posterior samples in Table~\ref{tab:jointttvrv} for these solutions. Interestingly, these solutions prefer larger mass ratios between planet b and candidate planet c than the 2:1 solution, and $<50\%$ out of the 200 randomly chosen posterior samples were found to be stable using the same criteria as adopted for the 2:1 solution.\footnote{We evaluated stability for smaller numbers of samples than we did for the 2:1 solution, because the purpose here is only to estimate the fraction of stable solutions.} As such, we deem these solutions less likely than the 2:1 case, although they are not fully excluded by the data. These solutions favor smaller masses for planet b than derived from the RV-only modeling, although the uncertainty is not well quantified.

In Figure~\ref{fig:ttvprediction}, we show TTV models for 20 stable posterior samples from each solution extended to the future. While the three solutions converge around the existing data by construction, they diverge occasionally, especially around the local minima of the periodic TTV curves. Thus the three solutions may be more conclusively distinguished with future transit timing data of TOI-2015b obtained at the right epoch. As such, we urge additional follow-up observations to help further constrain the potential period ratios in the system. The predicted transit times based on the 2:1 solution around this window are shown in Table~\ref{tab:predicted_transit_times}. The transits could happen $\sim 100\,\mathrm{min}$ later than these values if the 3:2 or 4:3 solution is correct; or they could happen earlier, if the true solution lies in a part of the parameter space we did not explore.

\newpage

\section{Discussion}
\label{sec:discussion}

\subsection{Mass and Bulk Composition}
\label{sec:mass-radius}
\begin{figure*}[htb!]
\centering
\includegraphics[width=0.7\textwidth]{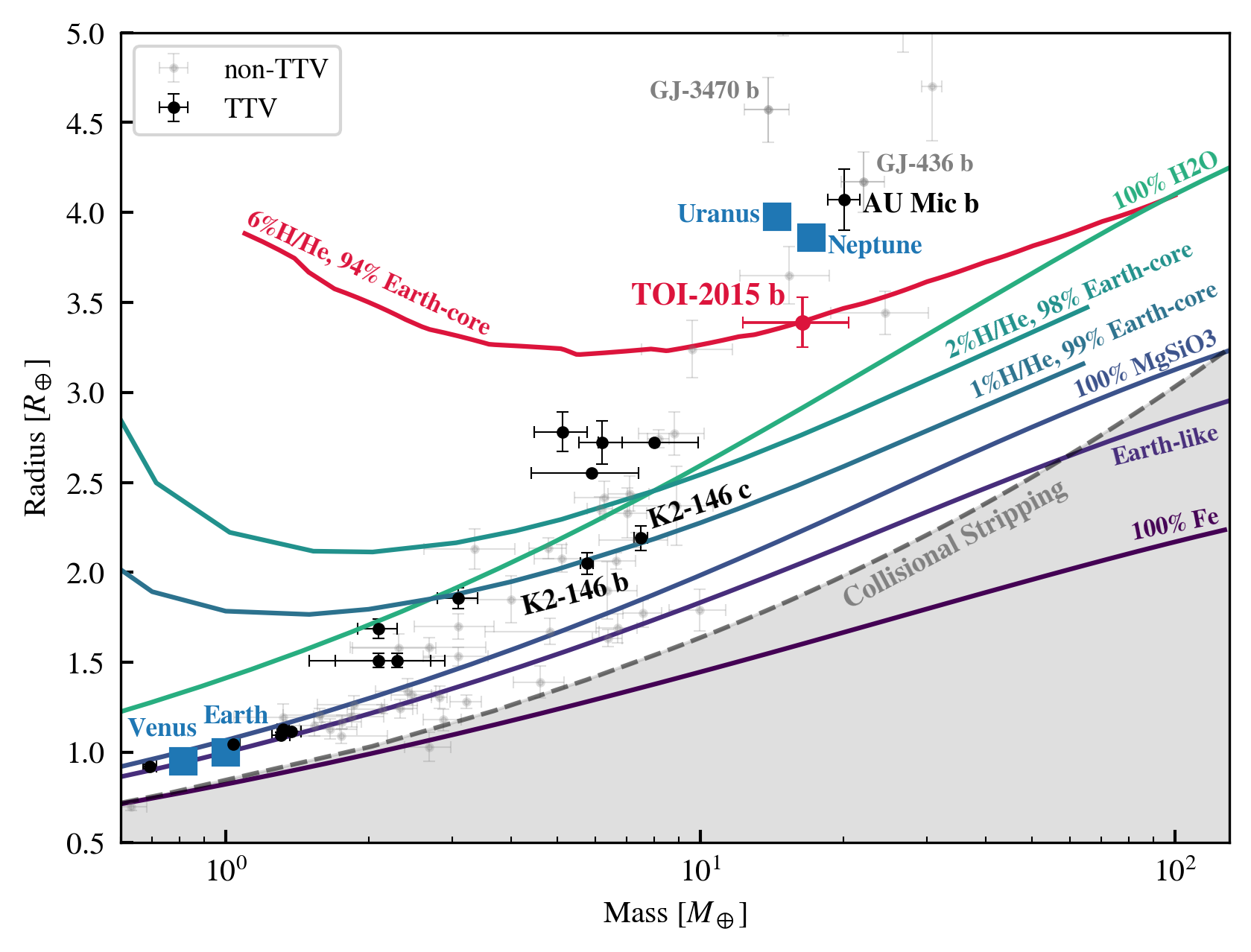}
\vspace{-0.4cm}
\caption{Constraints on mass and radius of TOI-2015b. M dwarf TTV planets are denoted with black points and non-TTV M dwarf planets are shown with faint grey points. Blue squares show solar system planets. TOI-2015b (red point) is similar to TTV planets AU Mic b and K2-146 b/c, which have potential composition models consistent with a rocky core and low mass H/He atmosphere. The red curve shows the two-layer composition model of \cite{lopez2014} with a rocky core shrouded by a H/He envelope, suggesting that TOI-2015b has a H/He mass fraction of \reshhe. The other solid lines show composition models from \cite{zeng2019}. The shaded grey region shows where the expected planetary iron content would exceed the maximum value predicted from models of collisional stripping from \cite{marcus2010}.}
\label{fig:mr}
\end{figure*}

Figure~\ref{fig:mr} shows the mass and radius constraints of TOI-2015b derived from our \texttt{juliet} joint photometry and RV fit. To gain an understanding of the possible composition of TOI-2015b consistent with its observed mass and radius, the plot displays lines of constant density from the models of \cite{zeng2019}. TOI-2015b's place among the density models suggests it is most likely volatile-rich.

To more precisely constrain the potential H/He mass fraction of TOI-2015b, we use the composition model from \cite{lopez2014}. This is a two-component model consisting of a rocky core enveloped by a H/He atmosphere, where the atmosphere is the dominant driver of the planet radius. One parameter the model relies on is age. In this study, we calculated SED and rotational-based age estimates for TOI-2015. Interpolating the tables from \cite{lopez2014} with the methodology in \cite{stefansson2020b}, we present mass fraction estimates for both age cases. When we assume the broad SED constraint of $6.8^{+4.7}_{-4.6}\unit{Gyr}$, we estimate a H/He mass fraction of 6.7\%. Using the gyrochronological age estimate of $1.1\pm 0.1 \unit{Gyr}$, TOI-2015b's mass and radius posteriors are compatible with a 6.0\% H/He envelope. The planet's mass fraction is not largely sensitive to the difference in age. This supports the \cite{lopez2014} models' demonstration of minimal contraction and change in the mass fraction for planets of age $\sim1\:$Gyr and older due to rapid cooling occurring at very early ages ($\sim100\:$Myr). Just as we adopt the gyrochronological age to be the primary age estimate for TOI-2015b, we too adopt a 6.0\% H/He envelope as the mass fraction. Figure~\ref{fig:mr} shows the 50th percentile model that best fits the observed mass and radius of TOI-2015b with the bold red line.

When we compare TOI-2015b to other planets of similar mass and size, the well-studied M dwarf planets GJ 436b and GJ 3470b are well-known targets with similar masses and radii to TOI-2015b, both of which are known to have volatile-rich compositions and evaporating atmospheres (see e.g., \citealt{bourrier2016} and \citealt{ninan2020}, respectively). In Figure \ref{fig:mr}, we also compare TOI-2015b to other M dwarf planets that show evidence of TTVs (black points). Those systems include AU Mic \citep{plavchan2020}, and K2-146 \citep{lam2020}, where TOI-2015b has a similar mass and radius to AU Mic b.

Constraining planet masses via radial velocity observations for low mass planets orbiting active stars can be challenging and observationally expensive, where TTVs provide a way to measure the mass independently of radial velocities. This is particularly advantageous for planets around active stars, where stellar activity signatures in radial velocities can dominate over planetary signatures. As Figure \ref{fig:mr} highlights, the number of Neptune-sized planetary systems with precise mass measurements is limited, and detecting and characterizing additional such systems can further aid in understanding the possible bulk and atmospheric compositions of such systems.

\begin{figure*}[htb!]
\centering
\includegraphics[width=0.97\textwidth]{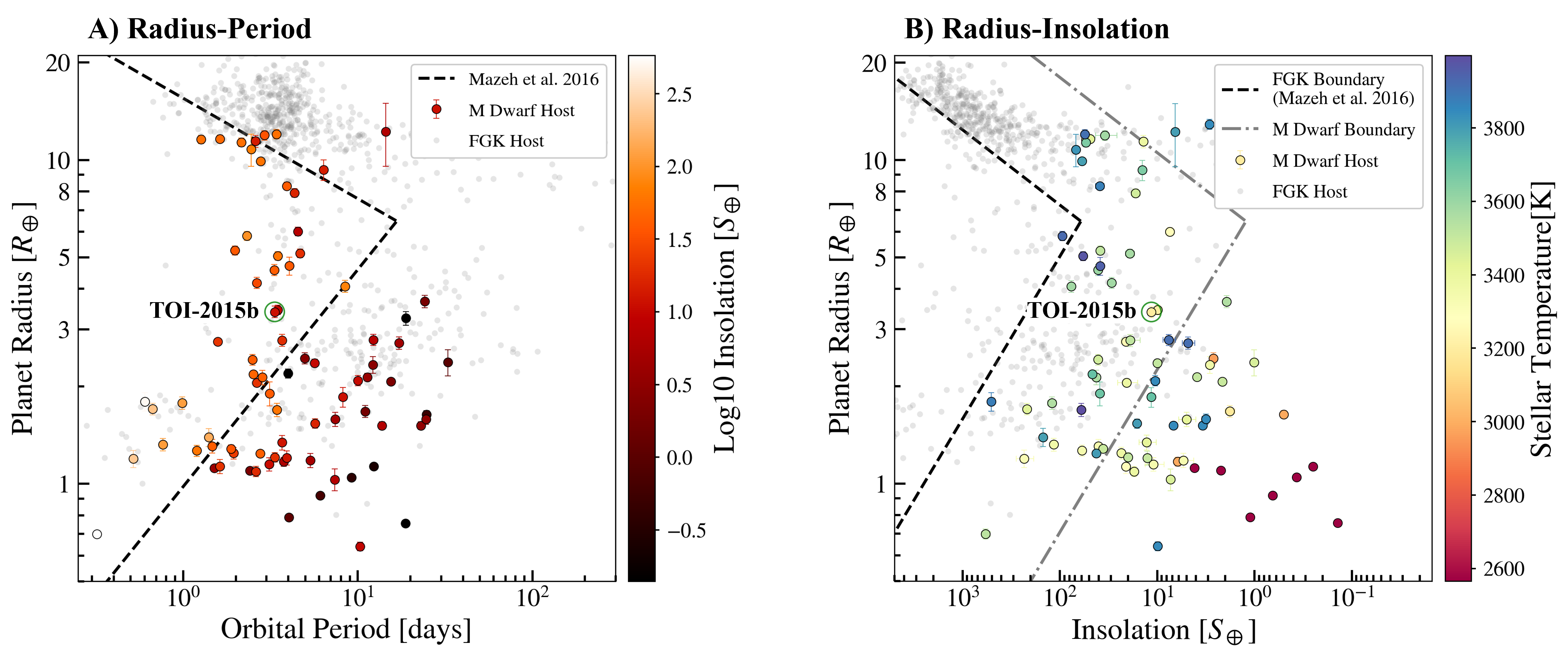}
\caption{The Neptune desert plotted in radius-period and radius-insolation parameter spaces. Colored points are M dwarf planets, and translucent grey points are planets with FGK hosts. TOI-2015b is circled in green. \textit{A}) Planet radius vs orbital period plane. Points are colored by Log10 insolation. The dashed line represents the boundary of the Neptune desert in the radius-period plane as defined by \cite{mazeh16}. \textit{B}) Planet radius vs insolation plane. Points are colored by stellar effective temperature. The black dashed line represents the Neptune desert boundary as defined by \cite{mazeh16}, assuming a Solar type star ($M_* = 1 M_\odot, L_* = 1 L_\odot$). The grey dot-dashed line is a redefined boundary assuming a M4 type star ($M_* = 0.35 M_\odot, L_* = 0.01 L_\odot$). While TOI-2015b falls within the \cite{mazeh16} Neptune desert boundary in radius-period space, its host star is a faint red mid M dwarf, so the planet receives too little flux to place it inside the Neptune desert.}
\label{fig:p-rad}
\end{figure*}

\subsubsection{Mass Loss and Atmosphere Retention}
Given a mass fraction of $\sim6\%$ for the H/He envelope, it is useful to estimate if TOI-2015b could retain its atmosphere as the system evolves. The atmospheres of close-in planets may experience mass loss due to extreme ultraviolet (EUV) irradiance from their host stars. The intense radiation in the upper atmosphere can facilitate the creation of hydrodynamic winds, resulting in atmospheric escape \citep{murray-clay2009, salz2016, debrecht2019}. To model this photoevaporative process and quantify the potential mass loss TOI-2015b may experience, we use the following energy-limited mass loss rate equation from \cite{murray-clay2009}:

\begin{eqnarray}
\dot{M}_{e-lim} \sim 6 \times 10^9 \left(\frac{\epsilon}{0.3}\right) \left(\frac{R_p}{10^{10}\: \textrm{cm}}\right)^3 \nonumber\\
\left(\frac{0.7 M_J}{M_p}\right) \left(\frac{F_{EUV}}{450\: \textrm{erg} \: \textrm{cm}^{-2} \: \textrm{s}^{-1}}\right) \textrm{g} \: \textrm{s}^{-1}
\end{eqnarray}

Here, the \textit{energy-limited} mass loss, $\dot{M}_{e-lim}$, means that essentially all UV radiation received by the planet is used to power photoevaporative winds. $\: \epsilon$ is a heating efficiency factor \citep[see][]{shematovich2014}. $R_p$ and $M_p$ are planet radius and mass respectively. And $F_{EUV}$ is the total EUV radiation flux of the host star. To estimate $F_{EUV}$ for TOI-2015, we integrated the Mega-MUSCLES spectrum of GJ 1132, a star that shares TOI-2015's M4 spectral type \citep{mega-muscles2021}. We specifically calculated the flux in the $300 - 911.6$\AA $\:$ range of the Mega-MUSCLES EUV spectrum, which is based on an empirical scaling relation using Ly$\alpha$ flux \citep{linsky2014}.

Assuming values of $\epsilon = 0.3$ \citep[see][]{murray-clay2009} and $F_{EUV} \sim 21\: \textrm{erg} \: \textrm{cm}^{-2} \: \textrm{s}^{-1}$ for TOI-2015, we estimate an energy limited mass loss rate of $\dot{M}_{e-lim}\sim6.3 \times 10^7\: \mathrm{g\:s^{-1}}$ or $\sim0.002\%\: \mathrm{Gyr^{-1}}$ given TOI-2015b's current planetary mass and radius. Using either the gyrochronological age estimate of  $1.1\pm 0.1 \unit{Gyr}$ or the SED estimate of $6.8^{+4.7}_{-4.6}\unit{Gyr}$ for the TOI-2015 system, the associated H/He envelope mass fractions of 6.0\% and 6.7\% respectively should be retained with a constant mass loss rate of $\dot{M}_{e-lim}\sim0.002\%\: \mathrm{Gyr^{-1}}$. With this estimate, however, we acknowledge that planets do not experience mass loss at a constant rate over their lifetimes; most mass loss occurs within the first 100 Myr \citep{owen2013}.

\subsection{Neptune Desert}
\label{neptune_desert}

Figure~\ref{fig:p-rad} displays M dwarf planets in relation to the ``Neptune desert'', a region around stars where a significant dearth of short-period (P < $\sim$4 days) Neptune-sized planets has been observed \citep{mazeh16}. Although the exact causes for the Neptune desert are still debated, the desert could be a relic of orbital migration combined with atmospheric evaporation, where the small, short-period planets we see in the desert were once the cores of Neptunes whose atmospheres were blown away by the intense radiation of their host stars \citep{owen2018}.

The Neptune desert is commonly discussed in terms of radius and period. However, recent work has highlighted the role of insolation in shaping the desert and how it varies as a function of stellar type \citep{mcdonald2019,kanodia2020,kanodia2021,powers2023}. The radius-period plane boundaries of the Neptune desert in \cite{mazeh16} were defined from observations primarily of FGK stars. To fully understand the landscape of Neptune-sized planets, we must encompass a wider variety of stellar types, including cooler host stars. In radius-insolation space, the boundaries of the desert region are redefined with host star type, taking into account how insolation can vary significantly between planets of similar orbital periods but different spectral classifications \citep{mcdonald2019}. 

Following \cite{kanodia2021}, we assess TOI-2015b in relation to the Neptune desert for both period and insolation parameters, plotted in Figure~\ref{fig:p-rad}. TOI-2015b falls within the Neptune desert boundary as defined by \cite{mazeh16} in period-radius space. However, if we adapt the period boundary into insolation space for FGK stars (which formed the bulk of the host stars in the Mazeh 2016 sample), then TOI-2015b falls outside of the Neptune desert (dashed line in Figure \ref{fig:p-rad}B). When we redefine the boundary using the mass and luminosity parameters of an M dwarf host, we see a less definitive dearth of planets in the region. As such, it is important that we consider radius-insolation space in our study of the ``Neptune desert'' for different types of host stars.

\subsection{Prospects for Rossiter McLaughlin Observations}
\label{sec:rm}
To date, most measurements of stellar obliquity---the angle between the stellar rotation axis and the planet's orbital axis---have been performed for Hot Jupiter (HJ) systems, which show a diversity of obliquities from well-aligned to highly misaligned orbits \citep{albrecht2012,albrecht2022}. This has been interpreted as signatures of how HJs form and evolve \citep{winn2015,albrecht2022}. However, because few small planets---which make up the bulk of known planetary systems---have their obliquity measured, it is unclear if the diverse range of obliquities observed for HJs is tied to the formation process of HJ planets specifically, or if it is a more general outcome of planet formation. As such, increasing the number of smaller planets with obliquity measurements will help yield further clues. Among these, warm Neptunes are particularly interesting, as they orbit far away enough from their stars where tidal forces that could realign them are negligible, making them pristine probes of the original obliquity angle. Interestingly, the few warm Neptunes that have their obliquities constrained around cool stars are observed to show a bifurcation of obliquities \citep[see e.g.,][]{stefansson2022,albrecht2022} from well-aligned (e.g., K2-25b, \citealt{stefansson2020b}) to close to polar (e.g., HAT-P-11b, \citealt{winn2009}; GJ 436b, \citealt{bourrier2018}; WASP-107b \citealt{rubenzahl2021}; GJ 3470b \citealt{stefansson2022}). Further, there is a rising sample of compact multi-planet systems that tend to be well-aligned \citep[e.g.,][]{frazier2023, albrecht2022}.

The relatively rapid rotation period of the host star and the 1\% planet transit depth make TOI-2015b a promising target for Rossiter-McLaughlin observations \citep[RM;][]{rossiter1924,mclaughlin1924} to constrain the obliquity of the system. Using $v~\sin~i = 3.2\pm0.6~\unit{km\ s^{-1}}$ from Table \ref{tab:stellarparam} and the radius ratio and impact parameter from Table \ref{tab:julietfitparams}, we use Equation 40 in \cite{winn2010b} to estimate an RM effect amplitude of $\Delta V_{\mathrm{RM}}\sim13~\unit{m/s}$ for TOI-2015b. This RM amplitude is similar to the RV semi-amplitude and is within range of the high-precision RV spectrographs on large telescopes that are sensitive in the red-optical or near-infrared wavelengths such as the Infrared Doppler instrument \citep[IRD;][]{kotani2014}, HPF \citep{mahadevan2012,mahadevan2014}, MAROON-X \citep{seifahrt2016,seifahrt2020}, ESPRESSO \citep{pepe2014}, and the Keck Planet Finder \citep[KPF;][]{gibson2016}. From the trends observed among Neptunes and multi-planet systems so far, it is possible TOI-2015b could be well-aligned. Obtaining additional obliquities of Neptunes around cool stars, such as TOI-2015b, will help shed further light into the orbital architectures and the obliquity distribution of close-in Neptune and multi-planet systems.

\section{Summary}
\label{sec:summary}
We report on the discovery of the warm Neptune TOI-2015b orbiting a nearby ($d=47 \unit{pc}$) active M4 dwarf using photometry from the TESS spacecraft, as well as precise photometry from the ground, precise radial velocity observations, and high-contrast imaging. The young host star is active and has a rotation period of $P_{\mathrm{rot}} = 8.7\pm0.9 \unit{days}$, and gyrochronological age estimate of $1.1 \pm 0.1 \unit{Gyr}$. The planet has an orbital period of $P_{\mathrm{orb}} = 3.349 \unit{days}$, a radius of $R_p~=~3.37_{-0.20}^{+0.15}~R_\oplus$ and a mass of $m_p = 16.4\pm4.1M_\oplus$ from a joint fit of the available photometry and the radial velocities, resulting in a planet density of $\rho_p~=~2.32_{-0.37}^{+0.38}~\unit{g/cm^3}$. These values are compatible with a composition of a rocky planetary core enveloped by a 6\% H/He envelope by mass. 

The system shows clear transit timing variations with a super period $P_{\mathrm{sup}}~\approx~430~\unit{days}$ and amplitude of $\sim~100~\unit{minutes}$, which we attribute being due to gravitational interactions with an outer planet candidate in the system. We show that the two available sectors of TESS show no evidence of additional transiting planets in the system. We do not see significant evidence for additional planets in the available RVs.

As planets close to period commensurability often show TTVs, to arrive at a mass constraint leveraging both the available RV data and the TTV data, we jointly modeled the RVs and the TTVs assuming an outer planet candidate at three likely period commensurabilities of period ratios of 2:1, 3:2, and 4:3 compared to TOI-2015b. We show that the system is well-described in the 2:1 resonance case, which suggests that TOI-2015b has a mass of $m_b = 13.3_{-4.5}^{+4.7} M_\oplus$---in agreement with the joint transit and RV fit---and that the outer planet candidate has a mass of $m_c = 6.8_{-2.3}^{+3.5} M_\oplus$. However, given the data available, other two-planet solutions---including the 3:2 and 4:3 resonance scenarios---cannot be conclusively excluded.

There are a number of characteristics that make TOI-2015b an interesting target for future studies. TOI-2015b is one of the most massive known M dwarf TTV planets, along with AU Mic b. With its large Transmission Spectroscopy Metric (TSM) of $\sim84$ (estimated using the equations in \cite{kempton2018}), it is an opportune target for atmospheric observations. This planet is also amenable to constrain obliquity with the Rossiter-McLaughlin effect, which would yield further insights into the possible formation history of the system. Most critically, we urge additional transit observations to help further understand the TTVs in the system and characterize the second planet candidate. To aid in future transit observations, we provide predictions of future transit midpoints assuming the 2:1 model is correct in Table \ref{tab:predicted_transit_times} in the Appendix.

\acknowledgments

We thank the reviewer for the thoughtful comments, which helped improve the strength of the manuscript. GS acknowledges support provided by NASA through the NASA Hubble Fellowship grant HST-HF2-51519.001-A awarded by the Space Telescope Science Institute, which is operated by the Association of Universities for Research in Astronomy, Inc., for NASA, under contract NAS5-26555. This work was partially supported by funding from the Center for Exoplanets and Habitable Worlds. The Center for Exoplanets and Habitable Worlds is supported by the Pennsylvania State University, the Eberly College of Science, and the Pennsylvania Space Grant Consortium. This work was supported by NASA Headquarters under the NASA Earth and Space Science Fellowship Program through grants 80NSSC18K1114. We acknowledge support from NSF grants AST-1006676, AST-1126413, AST-1310875, AST-1310885, AST-1517592, AST-1907622, AST-1909506, AST-1910954, AST-2108569, AST-2108801, the NASA Astrobiology Institute (NAI; NNA09DA76A), and PSARC in our pursuit of precise radial velocities in the NIR. Computations for this research were performed on the Pennsylvania State University’s Institute for Computational \& Data Sciences (ICDS). C.I.C. acknowledges support through an appointment to the NASA Postdoctoral Program at the Goddard Space Flight Center administered by ORAU through a contract with NASA. The research was carried out, in part, at the Jet Propulsion Laboratory, California Institute of Technology, under a contract with the National Aeronautics and Space Administration (80NM0018D0004).

These results are based on observations obtained with the Habitable-zone Planet Finder Spectrograph on the Hobby-Eberly Telescope. We thank the Resident astronomers and Telescope Operators at the HET for the skillful execution of our observations of our observations with HPF. The Hobby-Eberly Telescope is a joint project of the University of Texas at Austin, the Pennsylvania State University, Ludwig-Maximilians-Universität München, and Georg-August Universität Gottingen. The HET is named in honor of its principal benefactors, William P. Hobby and Robert E. Eberly. The HET collaboration acknowledges the support and resources from the Texas Advanced Computing Center. This work makes use of observations from the Las Cumbres Observatory global telescope network. We acknowledge publicly available data from the LCOGT Key Project (KEY2020B-005), which we retrieved from the LCOGT archive.

Some of the observations in the paper made use of the NN-EXPLORE Exoplanet and Stellar Speckle Imager (NESSI). NESSI was funded by the NASA Exoplanet Exploration Program and the NASA Ames Research Center. NESSI was built at the Ames Research Center by Steve B. Howell, Nic Scott, Elliott P. Horch, and Emmett Quigley.

We acknowledge the use of public TOI Release data from pipelines at the TESS Science Office and at the TESS Science Processing Operations Center. This research has made use of the Exoplanet Follow-up Observation Program website, which is operated by the California Institute of Technology, under contract with the National Aeronautics and Space Administration under the Exoplanet Exploration Program. This paper includes data collected by the TESS mission, which are publicly available from the Multimission Archive for Space Telescopes (MAST). Support for MAST for non-HST data is provided by the NASA Office of Space Science via grant NNX09AF08G and by other grants and contracts. This research made use of the SIMBAD database, operated at CDS, Strasbourg, France. This research made use of \texttt{lightkurve}, a Python package for Kepler and TESS data analysis (Lightkurve Collaboration, 2018). This research made use of the NASA Exoplanet Archive, which is operated by the California Institute of Technology, under contract with the National Aeronautics and Space Administration under the Exoplanet Exploration Program. This work has made use of data from the European Space Agency (ESA) mission {\it Gaia} (\url{https://www.cosmos.esa.int/gaia}), processed by the {\it Gaia} Data Processing and Analysis Consortium (DPAC, \url{https://www.cosmos.esa.int/web/gaia/dpac/consortium}). Funding for the DPAC has been provided by national institutions, in particular the institutions participating in the {\it Gaia} Multilateral Agreement.

\facilities{TESS, Gaia, HET 10m/HPF, WIYN 3.5m/NESSI, Lick 3m/ShaneAO, LCOGT 1m, WIRO 2.3m, RBO 0.6m, ARC 3.5m} 
\software{\texttt{AstroImageJ} \citep{collins2017}, 
\texttt{astroplan} \citep{morris2018},
\texttt{astropy} \citep{astropy2013},
\texttt{astroquery} \citep{astroquery},
\texttt{barycorrpy} \citep{kanodia2018}, 
\texttt{batman} \citep{kreidberg2015},
\texttt{corner.py} \citep{dfm2016}, 
\texttt{celerite} \citep{Foreman-Mackey2017}, 
\texttt{dynesty} \citep{speagle2019}, 
\texttt{EXOFASTv2} \citep{eastman2017},
\texttt{GALPY} \citep{bovy2015}, 
\texttt{GNU Parallel} \citep{Tange2011}, 
\texttt{HxRGproc} \citep{ninan2018},
\texttt{iDiffuse} \citep{stefansson2018b},
\texttt{jnkepler} (Masuda 2023, in prep.),
\texttt{Jupyter} \citep{jupyter2016},
\texttt{juliet} \citep{espinoza2019},
\texttt{matplotlib} \citep{hunter2007},
\texttt{MultiNest} \citep{2009MNRAS.398.1601F, 2014A&A...564A.125B},
\texttt{numpy} \citep{vanderwalt2011},
\texttt{pandas} \citep{pandas2010},
\texttt{radvel} \citep{fulton2018},
\texttt{SERVAL} \citep{zechmeister2018},
\texttt{transitleastsquares} \citep{hippke2019},
\texttt{TTVFast} \citep{deck2014}.}

\bibliography{references,references_km, 2015}
\bibliographystyle{yahapj}

\appendix

\section{TESS Photometry and Future Transits}

\subsection{Individual TESS Transits}
\begin{figure}[H]
\centering
\includegraphics[width=0.9\columnwidth]{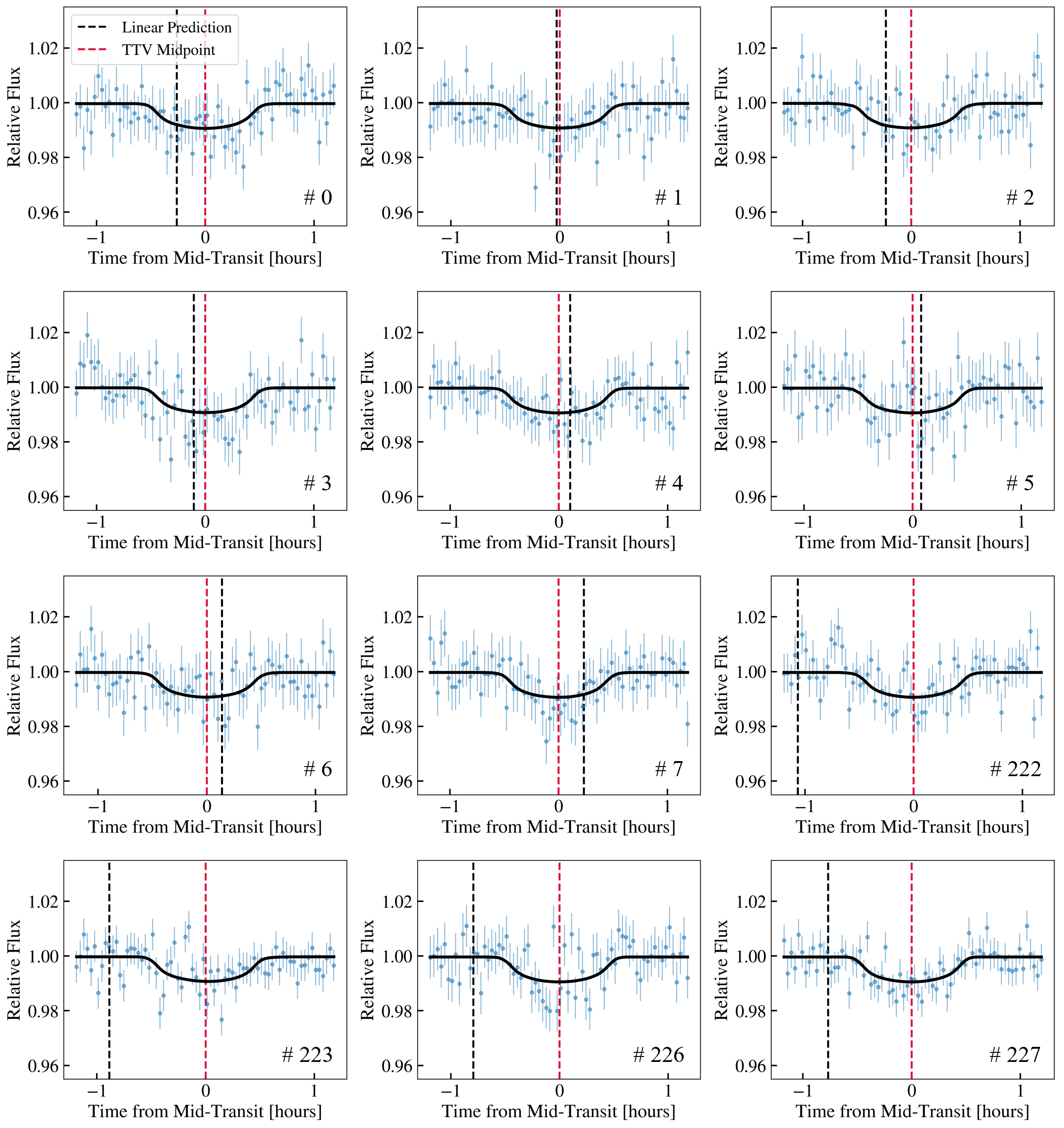}
\caption{Individual TESS transits detrended photometry with \texttt{juliet} TTV fit model. The vertical black lines represent the expected linear predictions, and the vertical red lines are the true TTV midpoints. Transit number is labeled in the bottom right corner of each transit.}
\vspace{-0.6cm}
\label{fig:tess_transits}
\end{figure}

Figure \ref{fig:tess_transits} shows each individual transit from TESS Sectors 24 and 51, accounting for the TTVs.

\subsection{TTVs}
Table \ref{tab:ttvs} shows the TTV midpoint time for each transit observation used in this study with their transit numbers from the first observed TESS transit (transit 0).

\begin{table}[H]
\centering
\caption{TTV midpoints from \texttt{juliet} joint fit. Instruments/Facilities: Transiting Exoplanet Survey Satellite (TESS), Las Cumbres Observatory Global Telescope (LCOGT), Wyoming Infrared Observatory (WIRO), Red Buttes Observatory (RBO), Astrophysical Research Council (ARC)}
\begin{tabular}{l c c c}
\hline\hline
Instrument &    Transit Number &    TTV (BJD)\\\hline
TESS    &   0   &   $2458956.0335^{+0.0025}_{-0.0026}$ \\
TESS    &   1   &   $2458959.3727^{+0.0038}_{-0.0033}$ \\
TESS    &   2   &   $2458962.7303^{+0.0039}_{-0.0049}$ \\
TESS    &   3   &   $2458966.0739^{+0.0029}_{-0.0026}$ \\
TESS    &   4   &   $2458969.4141^{+0.0023}_{-0.0023}$ \\
TESS    &   5   &   $2458972.7641^{+0.0026}_{-0.0026}$ \\
TESS    &   6   &   $2458976.1106^{+0.0034}_{-0.0039}$ \\
TESS    &   7   &   $2458979.4557^{+0.0028}_{-0.0026}$ \\
LCOGT   &   24   &   $2459036.3923^{+0.0011}_{-0.0011}$ \\
WIRO    &   137   &   $2459414.7537^{+0.0012}_{-0.0014}$ \\
TESS    &   222   &   $2459699.5379^{+0.0030}_{-0.0036}$ \\
TESS    &   223   &   $2459702.8793^{+0.0043}_{-0.0023}$ \\
TESS    &   226   &   $2459712.9224^{+0.0019}_{-0.0020}$ \\
TESS    &   227   &   $2459716.2703^{+0.0017}_{-0.0018}$ \\
RBO    &   329   &   $2460057.7924^{+0.0045}_{-0.0044}$ \\
ARC    &   329   &   $2460057.7974^{+0.00043}_{-0.00044}$ \\
ARC    &   332   &   $2460067.8488^{+0.00037}_{-0.00036}$ \\ \hline
\end{tabular}
\label{tab:ttvs}
\end{table}

\subsection{Future Predicted Transits}
Table \ref{tab:predicted_transit_times} lists future predicted transits assuming the near 2:1 resonance solution. 

\begin{table}[H]
\centering
\caption{Predicted future mid-transit times of TOI-2015b and their uncertainties for the 2:1 solution. The values shown are the mean and standard deviation calculated from the stable posterior samples derived in Section~\ref{sec:TTV}. We note that the transits could happen $\sim~100~\unit{min}$ later than these values if the 3:2 or 4:3 solution is correct. Only the first five entries are shown here; full table is available online.}
\begin{tabular}{ccc}
\hline\hline
Predicted transit midpoint (UTC) & Predicted transit midpoint (BJD) & Midpoint uncertainty (days)\\
\hline 
2024-06-01 10:09:07 & $2460462.923$ & $0.025$ \\
2024-06-04 18:36:00 & $2460466.275$ & $0.025$ \\
2024-06-08 03:02:53 & $2460469.627$ & $0.025$ \\
2024-06-11 11:28:19 & $2460472.978$ & $0.025$ \\
2024-06-14 19:55:12 & $2460476.330$ & $0.024$ \\
$\cdots$&$\cdots$ & $\cdots$ \\ 
\hline
\end{tabular}
\label{tab:predicted_transit_times}
\end{table}

\clearpage

\subsection{Corner plot of 2:1 solution}
\begin{figure*}[!htb]
\centering
\includegraphics[width=0.97\textwidth]{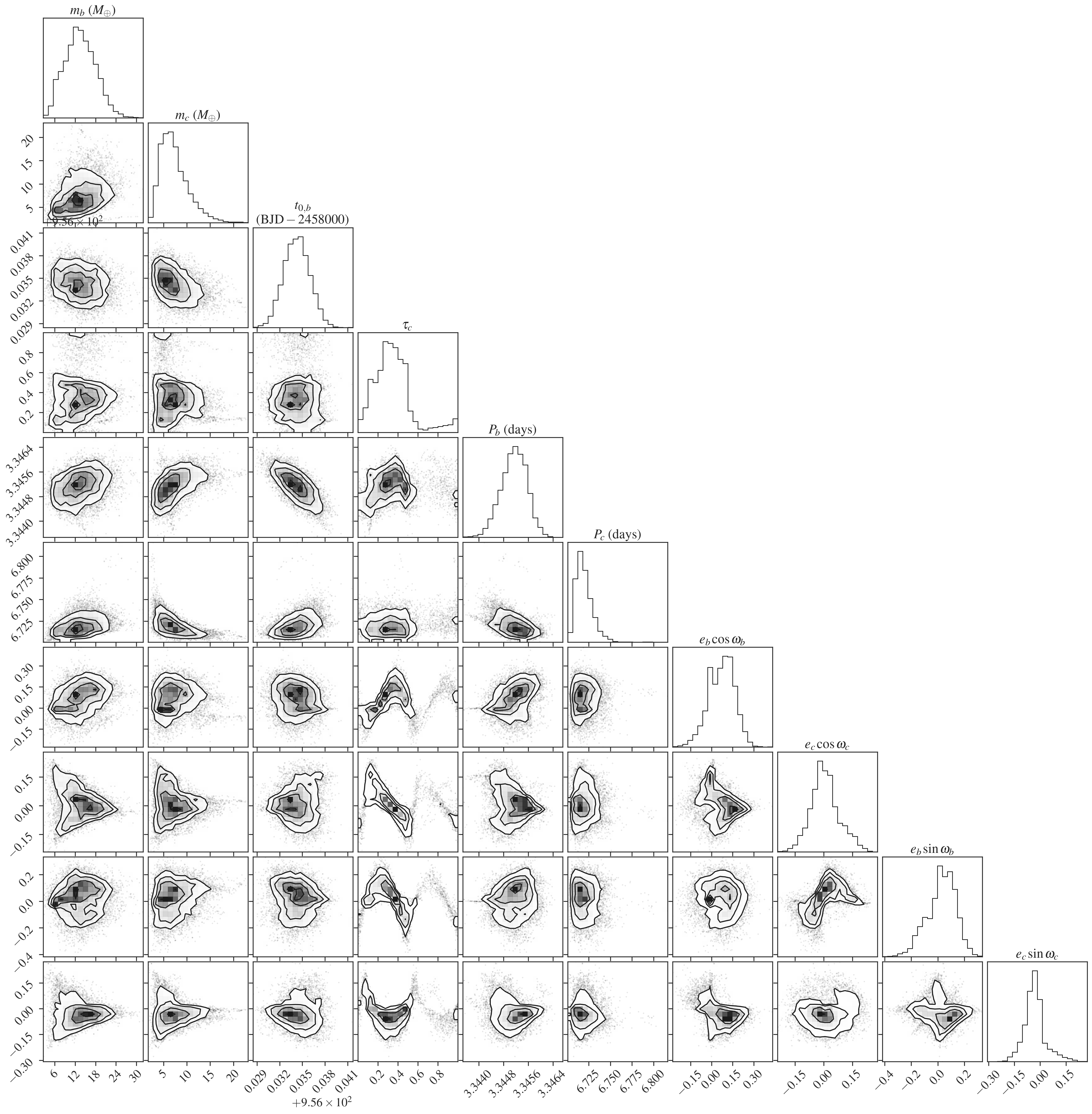}
\caption{Corner plot for the stable posterior samples from the joint TTV--RV fit assuming the 2:1 period ratio. The figure was created using {\tt corner.py} \citep{dfm2016}.}
\vspace{0.3cm}
\label{fig:ttv_corner}
\end{figure*}

Figure \ref{fig:ttv_corner} shows a corner plot of the posteriors from the 2:1 solution.

\clearpage

\newpage

\section{HPF RVs and Activity Indicators}
Table \ref{tab:rvs} lists the RVs from HPF and associated activity indicators derived from the HPF spectra used in this work.

\begin{table}[H]
\centering
\caption{HPF RVs used in this work along with the Differential Line Width (dLW), Chromatic Index (CRX), and the line indices for the three Ca II infrared triplet lines (Ca II IRT 1, 2 and 3), along with associated errors. These data are available as a machine readable table.}
\begin{tabular}{l c c c c c c}
\hline\hline
BJD &    RV [$\unit{m\:s^{-1}}$] &   dLW [$\unit{m^{2}\:s^{-2}}$]&   CRX  [$\unit{m\:s^{-1}\:Np^{-1}}$]&  Ca II IRT 1 &  Ca II IRT 2 &  Ca II IRT 3 \\ \hline
2459065.674261 &   $-6.9 \pm 9.2$ &    $63.3 \pm 24.3$ &    $61.3 \pm 99.1$ & $0.815 \pm 0.006$ & $0.548 \pm 0.006$ & $0.505 \pm 0.005$ \\
2459066.671994 &    $0.6 \pm 8.7$ &    $63.2 \pm 23.1$ &  $164.0 \pm 110.5$ & $0.832 \pm 0.006$ & $0.555 \pm 0.006$ & $0.513 \pm 0.005$ \\
2459085.624486 &   $15.0 \pm 9.1$ &    $38.6 \pm 24.4$ &  $202.5 \pm 114.3$ & $0.856 \pm 0.006$ & $0.547 \pm 0.006$ & $0.510 \pm 0.005$ \\
2459089.613160 &  $23.5 \pm 11.3$ &   $-18.1 \pm 30.4$ &   $48.2 \pm 118.3$ & $\ldots$          & $0.629 \pm 0.009$ & $0.584 \pm 0.007$ \\
2459096.599695 &  $-5.7 \pm 11.7$ &    $58.1 \pm 31.4$ &  $-75.4 \pm 157.7$ & $0.852 \pm 0.008$ & $0.570 \pm 0.009$ & $0.516 \pm 0.007$ \\
2459097.596633 &  $-19.4 \pm 9.1$ &    $36.0 \pm 24.5$ &  $-134.3 \pm 88.7$ & $1.105 \pm 0.007$ & $0.781 \pm 0.007$ & $0.695 \pm 0.006$ \\
2459218.030657 & $-17.6 \pm 24.7$ &   $-60.5 \pm 67.3$ &  $555.4 \pm 235.5$ & $0.932 \pm 0.021$ & $0.629 \pm 0.021$ & $0.539 \pm 0.019$ \\
2459219.032663 &   $4.1 \pm 20.1$ &    $22.3 \pm 54.2$ &  $199.1 \pm 243.6$ & $0.905 \pm 0.016$ & $0.621 \pm 0.017$ & $0.557 \pm 0.014$ \\
2459239.973718 & $-20.9 \pm 20.0$ &  $-148.7 \pm 54.8$ &   $89.4 \pm 248.5$ & $0.818 \pm 0.016$ & $0.517 \pm 0.017$ & $0.514 \pm 0.014$ \\
2459241.967614 &  $-6.1 \pm 13.9$ &    $32.0 \pm 37.5$ & $-264.6 \pm 227.0$ & $0.793 \pm 0.010$ & $0.522 \pm 0.010$ & $0.491 \pm 0.008$ \\
2459267.907011 &  $-18.8 \pm 8.9$ &    $15.2 \pm 23.9$ &  $-86.9 \pm 121.1$ & $0.843 \pm 0.006$ & $0.541 \pm 0.006$ & $0.497 \pm 0.005$ \\
2459271.881308 &  $-9.3 \pm 15.6$ &    $71.4 \pm 41.8$ &  $-21.5 \pm 194.2$ & $0.968 \pm 0.012$ & $0.638 \pm 0.011$ & $0.597 \pm 0.010$ \\
2459273.885647 & $-17.0 \pm 36.5$ & $-160.4 \pm 100.3$ & $-241.4 \pm 362.9$ & $0.838 \pm 0.027$ & $0.503 \pm 0.030$ & $0.418 \pm 0.027$ \\
2459275.875276 & $-33.2 \pm 13.6$ &   $-59.7 \pm 36.9$ &  $-91.5 \pm 111.3$ & $0.785 \pm 0.010$ & $0.519 \pm 0.010$ & $0.496 \pm 0.008$ \\
2459276.877445 &  $29.4 \pm 16.8$ &   $-16.9 \pm 45.5$ &  $315.8 \pm 173.6$ & $\ldots$          & $0.524 \pm 0.012$ & $0.506 \pm 0.010$ \\
2459278.870972 &  $-1.2 \pm 21.0$ &    $16.9 \pm 56.7$ &  $-53.6 \pm 238.2$ & $0.836 \pm 0.015$ & $0.538 \pm 0.017$ & $0.515 \pm 0.015$ \\
2459297.817717 &   $5.7 \pm 15.1$ &    $85.2 \pm 40.4$ & $-100.2 \pm 227.0$ & $0.907 \pm 0.012$ & $0.577 \pm 0.012$ & $0.520 \pm 0.010$ \\
2459324.975199 &  $-19.0 \pm 9.8$ &    $90.9 \pm 26.1$ & $-139.2 \pm 129.2$ & $0.871 \pm 0.007$ & $0.596 \pm 0.007$ & $0.555 \pm 0.005$ \\
2459325.963842 &   $6.8 \pm 31.8$ &    $96.6 \pm 84.2$ & $-485.4 \pm 277.7$ & $0.854 \pm 0.025$ & $0.684 \pm 0.025$ & $0.579 \pm 0.020$ \\
2459329.735950 &  $19.9 \pm 12.6$ &     $5.9 \pm 33.5$ &  $236.5 \pm 154.5$ & $0.805 \pm 0.009$ & $0.561 \pm 0.009$ & $0.508 \pm 0.007$ \\
2459329.958065 &  $31.5 \pm 14.4$ &    $74.0 \pm 38.2$ &   $10.5 \pm 162.5$ & $0.816 \pm 0.010$ & $0.583 \pm 0.011$ & $0.531 \pm 0.008$ \\
2459339.929512 &   $9.8 \pm 11.6$ &    $68.8 \pm 30.7$ &  $140.7 \pm 136.6$ & $0.850 \pm 0.008$ & $0.574 \pm 0.008$ & $0.536 \pm 0.006$ \\
2459351.683197 &  $-4.8 \pm 14.5$ &   $-50.3 \pm 39.3$ & $-157.3 \pm 177.3$ & $\ldots$          & $0.566 \pm 0.011$ & $0.506 \pm 0.008$ \\
2459359.659352 & $-21.3 \pm 14.3$ &   $-46.9 \pm 38.2$ &   $55.5 \pm 171.3$ & $0.886 \pm 0.010$ & $0.615 \pm 0.010$ & $0.561 \pm 0.008$ \\
2459360.866097 &  $11.9 \pm 15.6$ &    $33.5 \pm 41.6$ &  $342.8 \pm 220.5$ & $0.838 \pm 0.010$ & $0.574 \pm 0.011$ & $0.541 \pm 0.009$ \\
2459361.873109 &   $3.5 \pm 15.6$ &   $-52.2 \pm 41.8$ &   $36.0 \pm 101.3$ & $0.849 \pm 0.011$ & $0.560 \pm 0.011$ & $0.490 \pm 0.009$ \\
2459384.805438 &  $23.8 \pm 18.3$ &    $96.1 \pm 48.7$ & $-374.0 \pm 305.1$ & $0.979 \pm 0.013$ & $0.602 \pm 0.014$ & $0.582 \pm 0.011$ \\
2459388.793553 & $-28.1 \pm 16.7$ &   $-27.0 \pm 45.2$ &  $201.6 \pm 223.2$ & $0.955 \pm 0.012$ & $0.647 \pm 0.013$ & $0.537 \pm 0.010$ \\
2459414.718289 &   $5.6 \pm 18.3$ &   $-85.2 \pm 49.3$ & $-151.4 \pm 167.7$ & $0.895 \pm 0.012$ & $0.566 \pm 0.013$ & $0.523 \pm 0.010$ \\
2459416.714349 &  $21.2 \pm 12.5$ &   $-59.4 \pm 33.5$ &   $31.9 \pm 151.2$ & $0.914 \pm 0.008$ & $0.598 \pm 0.008$ & $0.560 \pm 0.007$ \\
2459450.621745 &  $15.0 \pm 13.1$ &     $4.6 \pm 35.1$ &   $-14.3 \pm 92.4$ & $0.908 \pm 0.009$ & $0.574 \pm 0.009$ & $0.584 \pm 0.007$ \\
2459597.992508 &  $11.9 \pm 19.9$ &   $-20.4 \pm 53.3$ &   $80.8 \pm 230.4$ & $0.862 \pm 0.013$ & $0.559 \pm 0.013$ & $0.522 \pm 0.011$ \\
2459605.972370 &  $25.8 \pm 10.8$ &   $110.7 \pm 28.8$ &   $44.3 \pm 124.4$ & $\ldots$          & $0.542 \pm 0.007$ & $0.524 \pm 0.006$ \\
2459626.920928 &   $9.6 \pm 14.9$ &    $14.7 \pm 40.2$ & $-151.6 \pm 214.0$ & $0.797 \pm 0.009$ & $0.496 \pm 0.009$ & $0.481 \pm 0.009$ \\
2459629.913211 &  $14.8 \pm 12.3$ &   $-56.2 \pm 33.4$ &  $-97.5 \pm 147.7$ & $0.921 \pm 0.008$ & $0.579 \pm 0.008$ & $0.567 \pm 0.007$ \\
2459635.896290 &  $34.3 \pm 14.8$ &  $-118.8 \pm 40.2$ &   $-1.8 \pm 154.9$ & $0.840 \pm 0.010$ & $0.544 \pm 0.011$ & $0.515 \pm 0.009$ \\
2459704.703361 & $-17.5 \pm 14.1$ &   $281.2 \pm 37.0$ &   $53.4 \pm 167.1$ & $0.863 \pm 0.010$ & $0.572 \pm 0.009$ & $0.550 \pm 0.008$ \\
\hline
\end{tabular}
\label{tab:rvs}
\end{table}

\end{document}